\documentclass[]{aastex631}

\begin{document}

   \title{A study of the star clusters' population in the giant molecular cloud G174+2.5}

\author[0000-0001-7039-7670]{Tatyana A. Permyakova}
\affiliation{Ural Federal University,
51 Lenin Street, Ekaterinburg, 620000, Russia}

\author[0000-0002-0155-9434]{Giovanni Carraro}
\affiliation{Dipartimento di Fisica e Astronomia,
Universita' di Padova,
Vicolo Osservatorio 3, I-35122, Padova, Italy}

\author[0000-0001-8669-803X]{Anton F. Seleznev}
\affiliation{Ural Federal University,
51 Lenin Street, Ekaterinburg, 620000, Russia}

\author[0000-0001-7575-5254]{Andrej M. Sobolev}
\affiliation{Ural Federal University,
51 Lenin Street, Ekaterinburg, 620000, Russia}

\author[0000-0002-3773-7116]{Dmitry A. Ladeyschikov}
\affiliation{Ural Federal University,
51 Lenin Street, Ekaterinburg, 620000, Russia}

\author[0000-0003-4338-9055]{Maria S. Kirsanova}
\affiliation{Institute of Astronomy of the Russian Academy of Sciences, 48 Pyatnitskaya Street, Moscow, 119017, Russia}
\affiliation{Ural Federal University,
51 Lenin Street, Ekaterinburg, 620000, Russia}

\begin{abstract}
   We study the structure, interstellar absorption, color-magnitude diagrams, kinematics, and dynamical state of embedded star clusters in the star-forming region associated with the giant molecular cloud G174+2.5.
   Our investigation is based on photometric data from the UKIDSS Galactic Plane Survey catalog and astrometric data from the Gaia DR3 catalogs.
   First, we recover all the known embedded clusters and candidate clusters in the region using surface density maps.
   Then, for the detected clusters, we determine their general parameters: the center positions, radii, number of stars, and reddening.
   To evaluate the reddening, we use both the NICEST algorithm and the Q-method.
   Both methods produce consistent extinction maps in the regions of the four studied clusters. However, the Q-method yields a much smaller color scatter in the CMD.
   For four clusters in particular (S235~North-West, S235~A-B-C, S235~Central, and S235~East1+East2), we were able to compute individual membership probabilities, the cluster distances, the cluster masses, and their  average proper motions. By building on these results, we have
   studied the clusters' kinematics and dynamics .
   Moreover,  we  estimate the mass of the gas component and the star formation efficiency (SFE) in the regions of these four clusters.
   Finally, we provide an estimate of the total energy of the stellar and gas components in the area of  these four clusters to determine whether the clusters are bound (here we consider a `cluster' as the system `stars + gas').
   The gravitational bound strongly depends on the region for which we estimate the gas mass.
   If we consider the mass of the entire cloud, all these four clusters turn out to be bound.
\end{abstract}


%

\section{Introduction}\label{1}

   Young star clusters are among the main sources of information about the process, rate, and efficiency of star formation.
   Their study allows one to establish the form of the initial mass function, to probe the position of the spiral arms of the Galaxy and to investigate the gravitational interaction of newly born stars and their interaction with the parental gas cloud on a variety of scales. In general, they are excellent benchmarks of the theory of star formation \citep{Lada03,GCa2020}.
   The employment of thermal imaging cameras and spectrometers with optical and infrared-optimized telescopes over the past decades \citep[for example,][]{Skrutskie06, Lawrence07, Werner04} had made it possible to study systematically the clusters' emergence inside molecular clouds.
   Clusters in this early evolutionary stage are often referred to as embedded star clusters \citep{Stahler18}.
   Much work has been done over the years to unravel the properties of these objects \citep{Lada03, Sobolev17, Stahler18}.
   Perhaps, the most important outcome is that most stars form in such proto-clusters, and that only a small fraction of these proto-clusters survive their emergence from molecular clouds and become long-lived open clusters.
   Currently, about 3000 embedded clusters and cluster candidates are known, and the corresponding information can be found in various studies and catalogs \citep[see, for example,][etc.]{Bica03, Lada03, Kumar06, Solin12, Camargo16, Ryu18}.

   In particular, the studies of \citet{Arzoumanian11} and \citet{Andre14} have shown that stars form in thin filaments of molecular gas.
   At the same time, observations indicate that dense gas in the star-forming molecular clouds \citep{Schisano14, Konyves15} has a multi-scale, fractal structure.
   At close distances, we can clearly distinguish the fine structure of clouds~-- thin filaments with stars forming in them.
   For more distant regions, on the contrary, small structures are not resolved well during observations, but large ones are visible.
   In this case, we cannot trace in which particular thin filaments individual stars are born, but it becomes clear that concentrations of young stars are located within large-scale structures.
   The origin of filaments and their general properties are still not completely understood, but filaments have become one of the main components of the modern star formation paradigm \citep{Andre14}. 
   For the purpose of this study, it is very important to underline that star clusters are found to form in the large-scale gaseous filaments.
   
   This study is devoted to the search and characterization of young star clusters  in one of the nearest star formation regions, which is associated with the giant molecular cloud (GMC) G174+2.5.
   In this region, \citet{Bieging16} indicate evidence for gas filaments of different scales.
   Besides, this area contains four regions of ionized hydrogen from the Sharpless catalog visible in the optical range: Sh2-231, Sh2-232, Sh2-233, and Sh2-235.
   The G174+2.5 complex is located in the Perseus spiral arm, and
   the estimate of its distance ranges from 1.4 to 2.5 kpc \citep{Georgelin73, Israel78, Evans81, Foster15}.
   The most valuable estimate of this complex distance by the trigonometric parallaxes of the maser sources was 1.56 kpc \citep{Burns15}.
   The HII region Sh2-235 is generated by the O9.5V  star BD$+35^{\circ}1201$ \citep{Georgelin73}, which ionizes and scatters the surrounding molecular gas.
   
   The Sh2-235 molecular complex is a region of ongoing star formation, where one can observe objects in different evolutionary phases.
   There are also young star clusters around Sh2-235.
   These clusters were detected from the analysis of Spitzer data in the vicinity of Sh2-235 \citep{Allen05, Chavarria14} and from 2MASS images in the direction of the IRAS source 05382+3547 in the same vicinity \citep{Kumar06}.
   Recently, the gas component of the region, young stellar objects and their spatial distribution have been studied in more details \citep[see, for example,][]{Porras+2000, Dewangan11, Dewangan16, Burns+2019PASJ}.
   \citet{Beuther+2002} and \citet{Ginsburg+2009} indicated that the gas dynamics around star cluster associated with the source IRAS 05358+3543 (star cluster 7 in our designation, see below) is determined by the outflows from young stellar objects.
   However, the characteristics of star clusters and their kinematics remain unstudied.
   
   The latest studies of the clusters in this region of star formation were carried out in \citet{Kirsanova08} and \citet{Camargo11}.
   \citet{Camargo11} estimate the fundamental parameters of extremely young open star clusters (OSCs) in this region, while
   \citet{Kirsanova08} discuss the possibility of sequential star formation in this region.
   The more recent work of \citet{Kirsanova14} continues the study of the mechanisms of star formation in the region.
   They conclude that the clusters S235~East 1, S235~East 2, and S235~Central most likely formed as a result of induced star formation caused by the expansion of the HII region.
   On the other hand, the S235~A-B-C cluster is quite distant from the expanding HII shell and therefore most likely has a spontaneous star formation mechanism \citep{Kirsanova08}.
   Such diversity in one region allows us to study how the characteristics of clusters differ depending on star formation mechanisms.

      \begin{figure}
      \centering
      \includegraphics[width=8cm]{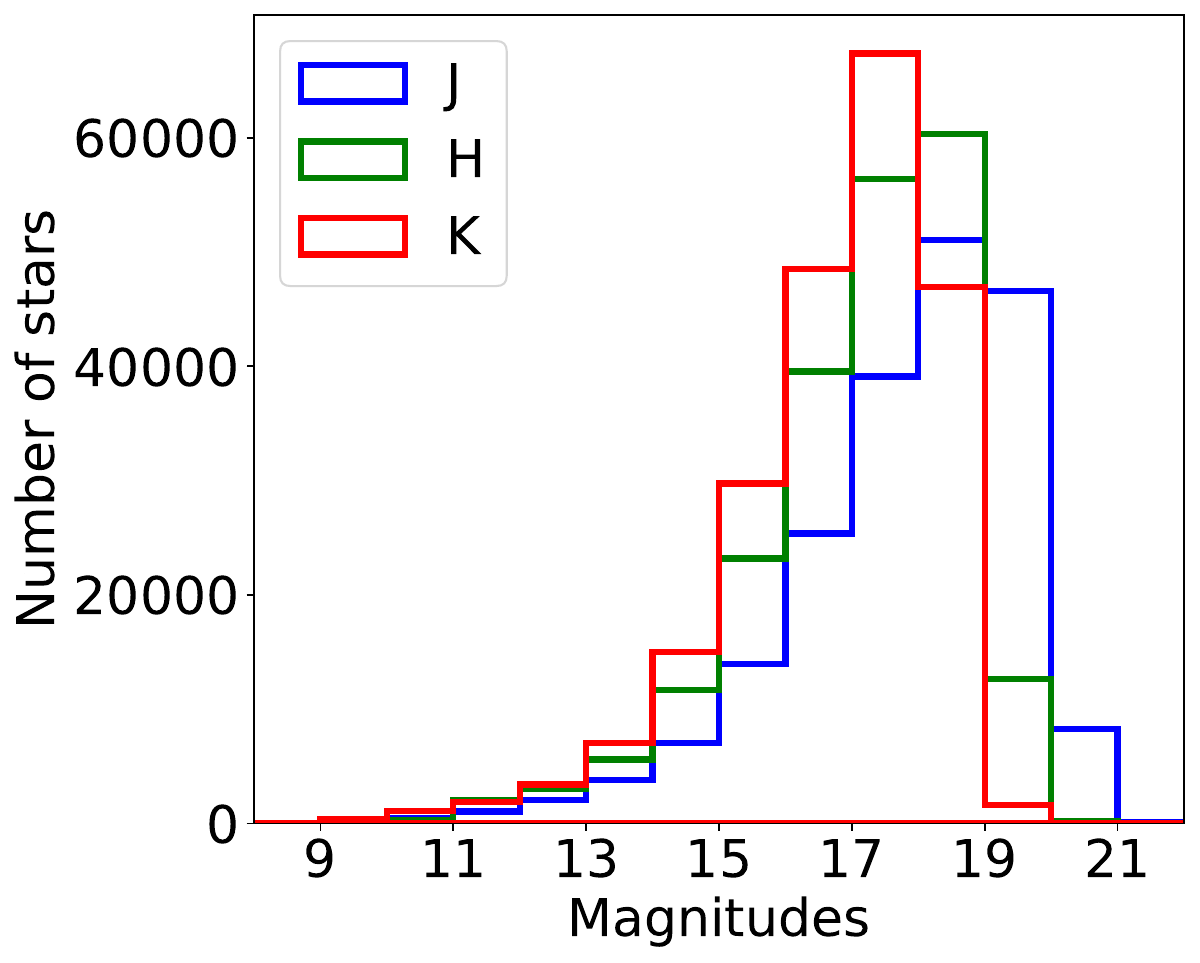}
      \caption{Histogram of the stellar magnitudes in the J, H, and K bands for a sample from the UKIDSS catalog in the G174+2.5 region.}
              \label{FigHistJHK}
    \end{figure}
    
   \begin{figure}
      \centering
      \includegraphics[width=17cm]{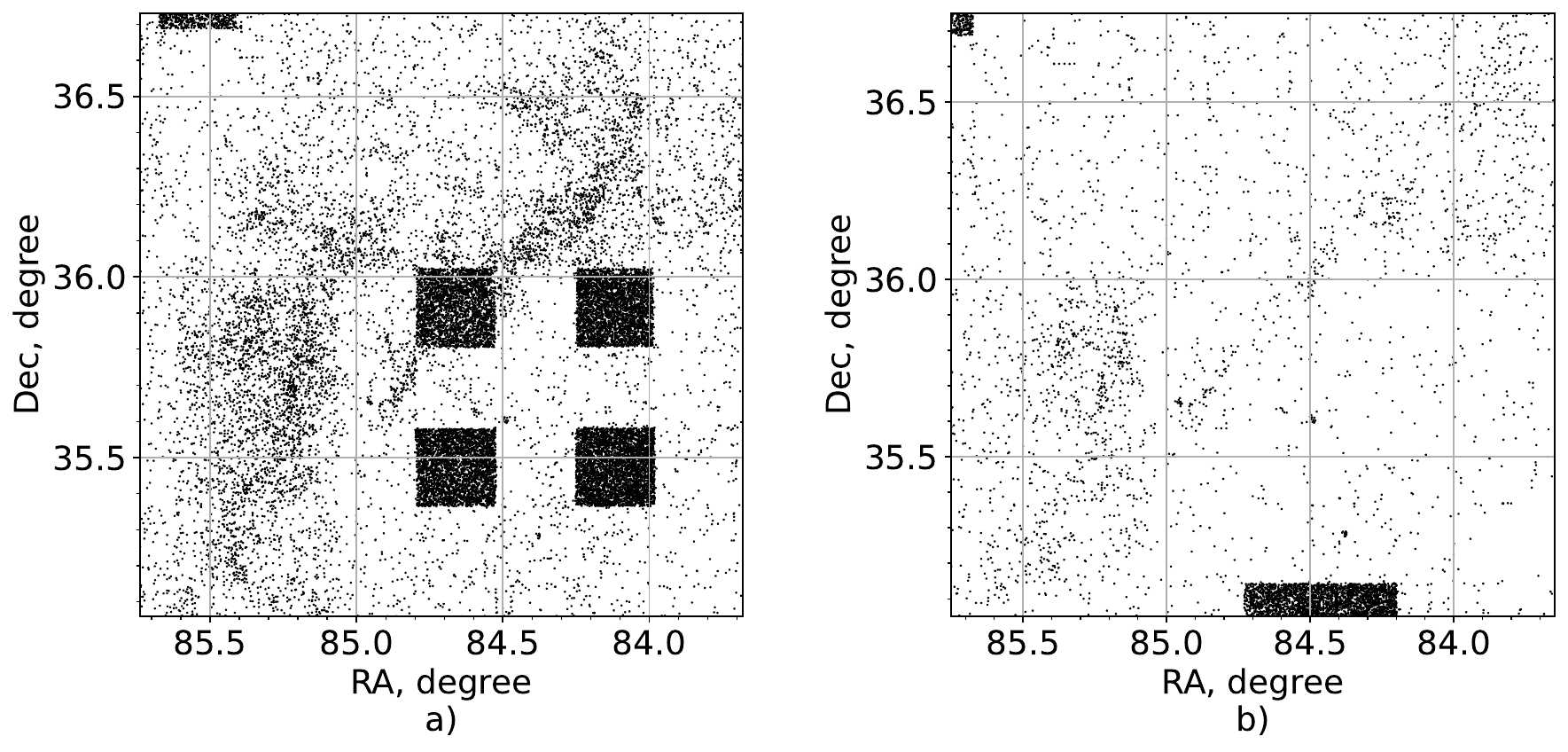}
      \caption{Star distribution maps for the UKIDSS sample of the G174+2.5 region; stars without magnitude data in the J (a) and H (b) bands.}
              \label{FigHJmis}%
    \end{figure}
   
   \citet{Burns+2019PASJ} marked that star clusters in the G174+2.5 complex are connected with the filaments twisted around Sh2-235 HII region.
   They proposed that clusters S235~East 1, S235~East 2, S235~Central, and S235~A-B-C could belong to the filament directed approximately along the line-of-sight.
   
   These previous studies were conducted with 2MASS data \citep{Skrutskie06}, which are shallow and have poor spatial resolution.
   To improve on this, we exploit much fainter and better resolution data from the UKIDSS catalog \citep{Lawrence07}.
   In particular, we employ data from the  UKIDSS Galactic Plane Survey (GPS), which covers almost the entire Milky Way band.
   For the purpose of studying the kinematics of stars in the clusters, we complement infrared photometry with the data from the Gaia DR3 catalog \citep{Gaia23}, which gives better accuracy of the proper motions and a larger sample of stars than previous Gaia releases.
   
   The paper is organized as follows.
   In Section \ref{2}, we present an overview of the catalog data used in the work.
   In Section \ref{3}, we present the NICEST absorption map in the region and discuss the selection of the filaments in our region.
   Section \ref{4} is devoted to the identification of clusters and the determination of their parameters.
   The kinematics of cluster stars are discussed in Section \ref{5}. 
   In Section \ref{7}, we estimate the mass of some clusters.
   In Section \ref{8}, we determine gas masses and star formation efficiencies in the regions of the four clusters. 
   Section \ref{9} includes an analysis of the possible future fate of some of the clusters under consideration.
   Finally, our conclusions are summarized in Section \ref{6}.

\section{UKIDSS and GAIA DR3 data}\label{2}

   This work uses data from the UKIDSS GPS catalog (version DR10PLUS) \citep{Lucas08} (hereinafter referred to as UKIDSS).
   The UKIDSS survey was carried out between 2005 and 2014 at the United Kingdom Infrared Telescope (UKIRT) in Hawaii, using the Wide Field Camera (WFCAM).
   All the details of this survey can be found in \citep{Casali07}.
   To study the G174+2.5 region, we took an area with the center at the point $RA=84.7^{\circ}$ and $Dec=35.9^{\circ}$ and a radius of 60 arcminutes.
   This area covers a number of HII regions and a large gas-dust filament.
   We consider the resulting sample as complete down to magnitudes $J \approx 18.0, H \approx 18.0$, and $K \approx 17.0$.
   These values are obtained from the magnitude  distribution of the stars  in the field, shown as a histogram in Fig.\ref{FigHistJHK}.
   We defined the limit of completeness as the magnitude beyond which the number of stars in the region decreases.
   The resulting limits are shallower compared to the reported limits for uncrowded fields in UKIDSS \citep{Lucas08}.
   The UKIDSS completeness is superior to 2MASS catalog, which can be considered as complete up to magnitudes $J \approx 15.8, H \approx 15.1$, and $K_S \approx 14.3$ \citep{Skrutskie06}.
   Due to this circumstance, we expect to obtain a more complete sample of low-mass stars, compared to previous studies.
   It is worth noting that when one takes into account stars with higher stellar magnitudes, 
   the results of star counts can give a different result comparing to the case of brighter stars.
   These differences may manifest themselves in the small-scale structure of clusters.
   Some overdensities can blend with the field or with neighbor clusters.
   Otherwise, the previously undetected overdensities can be found.
   
   However, the UKIDSS photometry of our region is characterized by uneven completeness of data in the J and H bands (see Fig.\ref{FigHJmis}).
   There are squared regions with a complete lack of data in these bands, that is related to the UKIRT sky coverage strategy.
   Concentrations of stars with missing photometry are also observed in regions with high absorption.
   This circumstance prevents to estimate the membership probability, absorption and mass of individual stars in these regions with our methods (Sections \ref{4.2}, \ref{7.1}, and Appendix A, since they require photometry data in all three bands.
   Nevertheless, the K band does not have `empty' fields.
   Then, we base our star counts and the surface density mapping on the K band which is generally uniform.
   If the cluster falls to the region with an absence of J or H photometry, we register it but can not perform further investigation.

   The catalog also contains several false detections.
   Some of them are located on the frame boundaries --- usually two or three stars with almost identical coordinates, some are concentrated around bright stars \citep{Solin12}.
   Due to their compact structure, such groups make a significant contribution to the surface density.
   Some of them look like real clusters, but in UKIDSS images, they appear as single bright stars.
   We find these false sources using the HDBSCAN \citep{Campello13} algorithm of the scikit-learn library \citep{Pedregosa11} of the Python language \citep{Pedregosa11}, using the dbscan\_clustering function.
   Clustering is carried out along equatorial coordinates.
   To find false sources at frame boundaries, we ran the program using the parameters min\_cluster\_size=2 and cut\_distance=0.6 arcsec (this value corresponds roughly to a 3-sigma chop given the typical positional errors in UKIDSS catalog astrometry).
   In this case, the search was conducted among all the stars in the region.
   We searched for groups of false sources concentrated around bright stars only among stars with K-band magnitudes less than $13^m$.
   In this case, we use the parameters min\_cluster\_size=3 and cut\_distance=5.0 arcsec.
   The latter value was chosen experimentally so that the algorithm would highlight the largest groups of false sources around bright stars.
   In each selected group, we retain the star with known photometry in the largest number of bands and remove the remaining stars from the sample.
   
   The original catalog data were supplemented with extinction values determined using the NICEST method \citep{Lombardi09} (for details, see Section \ref{3}) and rectangular x, y coordinates \citep[for more details see][]{Danilov20}.
   The x, y axes form a right-handed Cartesian coordinate system.
   The (x, y) plane coincides with the tangential plane.
   Rectangular coordinates were constructed so that their center coincides with the center of the region under consideration, the x and y axes are directed towards increasing RA and Dec, respectively.
   The projection of the celestial sphere onto a plane is necessary for plotting the surface density maps and radial density profiles using the KDE method (see Section \ref{4.1}).
   
   We also used the latest release of Gaia DR3 \citep{Gaia16,Gaia23} data to obtain information on the proper motions and the parallaxes of stars and \citet{Chavarria14} for the data on the positions and classes of the young stellar objects (YSO).
   The comparison of sources in the UKIDSS and Gaia DR3 catalogs, and the YSO list was carried out by the TOPCAT \citep{Taylor05} using the equatorial coordinates of the stars with the maximum error of 1 arcsec.

\section{NICEST absorption map and filament extraction}\label{3}

   \begin{figure}
      \centering
      $\vcenter{\hbox{\includegraphics[width=9.5cm]{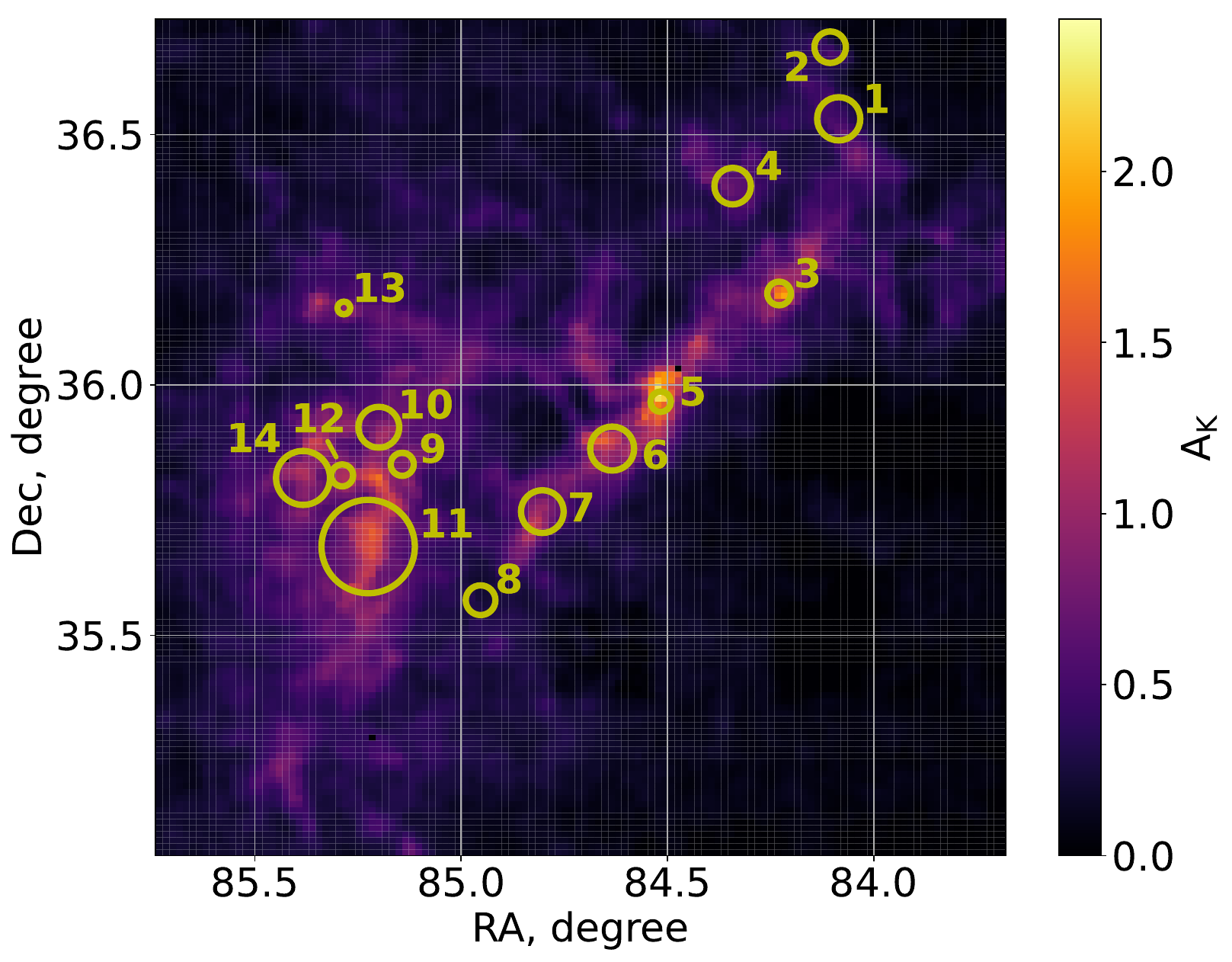}}}$
      \hspace{0.1cm}
      $\vcenter{\hbox{\includegraphics[width=8cm]{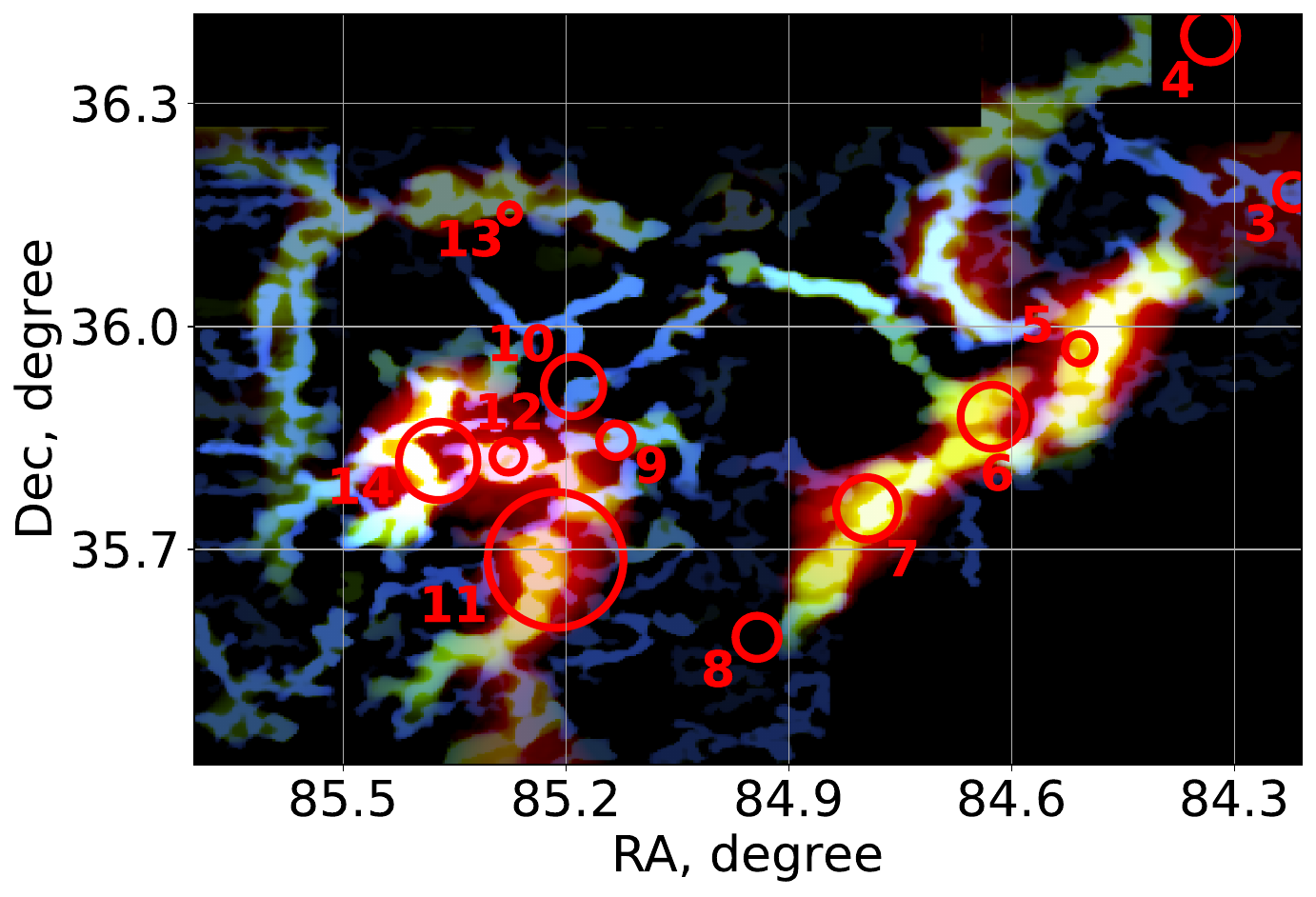}}}$
      \caption{The left panel shows the absorption distribution in the star forming region G174+2.5 obtained by the NICEST method.
      The yellow circles show the areas occupied by clusters, the numbers are the cluster IDs in this work.
      The right panel shows a map of the filaments identified in the star forming region G174+2.5 using the GetFilaments algorithm \citep{Menshchikov13}.
      Filaments of different scales are shown in different colors: blue -- 0.8 pc, green -- 3 pc, and red -- 12 pc. The scale of both panels is the same.}
      \label{FigExtMap}%
    \end{figure}

   We used data from the UKIRT deep sky infrared survey \citep{Lawrence07} together with data from the 2MASS survey \citep{Skrutskie06} to construct NICEST \citep{Lombardi09} extinction maps.
   The NICEST method is similar to conventional methods for determining interstellar  extinction in regions of low and moderate column density, but it has been specifically designed to search for small-scale structures of absorbing matter in regions of strong extinction.
   It is based on a comparison of the observed colors in the J, H and K bands and the stars’ absolute colors corresponding to zero absorption.

   We have constructed an absorption map for the G174+2.5 region with the center and radius indicated above.
   When constructing it, we did not consider UKIDSS catalog stars for which there was no photometric data in two bands, or for which the magnitude uncertainty was larger than $1^m$.
   Also due to the inaccuracy of the UKIDSS catalog for bright stars, we replaced the UKIDSS photometry with 2MASS for stars brighter than 12 mag in each band using the TOPCAT software \citep{Taylor05}.
   2MASS photometry was corrected to the UKIDSS photometric standard using equations (6)~(8) of \citet{Hodgkin09}.

   The extinction map of the star formation region G174+2.5 was obtained with a resolution (FWHM) of 1.0--1.5 arcmin.
   The pixel size was set to FWHM/3.
   The extinction map in the star-forming region G174+2.5 is shown in Fig.\ref{FigExtMap}, left panel.
   It is important to note that the NICEST method is a statistical method and does not provide the ability to determine extinction for individual stars.
   One can readily see that the high absorption regions follow the structure of  the molecular gas  filaments in the region around the known clusters in Fig. \ref{FigExtMap}, right panel.
   These filaments were extracted using the GetFilaments algorithm \citep{Menshchikov13} based on the map of the brightness temperature integrated over the Local Standard of Rest (LSR) velocity range $(-28,-2)$ km/s of the ${}^{13}CO(2-1)$ line in the star forming region G174+2.5 \citep{Bieging16}.
   
   The average extinctions for clusters with the individual membership probabilities (indicated in the paper as \textbf{10 (S235~North-West), 11 (S235~A-B-C), 12 (S235~Central), and 14 (S235~East1+East2)}) are given in Table \ref{TabClustPar}.

\section{Study of clusters}\label{4}

\subsection{Surface density maps}\label{4.1}

   We constructed surface density maps to identify areas, potentially containing previously undiscovered clusters.
   The maps were constructed using stars with interstellar extinction-corrected K-band magnitudes: $K_0 = K - A_K$.
   We chose this bandwidth because in other bands (J and H) the catalog coverage is not uniform (see Section \ref{2}), which would affect the detection of clusters.
   The value of the total extinction in the K band for every star was taken directly from the extinction map (see Section \ref{3}) for the coordinates of the star.
   Correcting the magnitudes for interstellar extinction increases the contrast of the stellar density map, which is important for star-forming regions, where  differential extinction is common.
   
\begin{figure}
      \centering
      \begin{minipage}[h]{0.49\linewidth}
      \center{\includegraphics[width=1\linewidth]{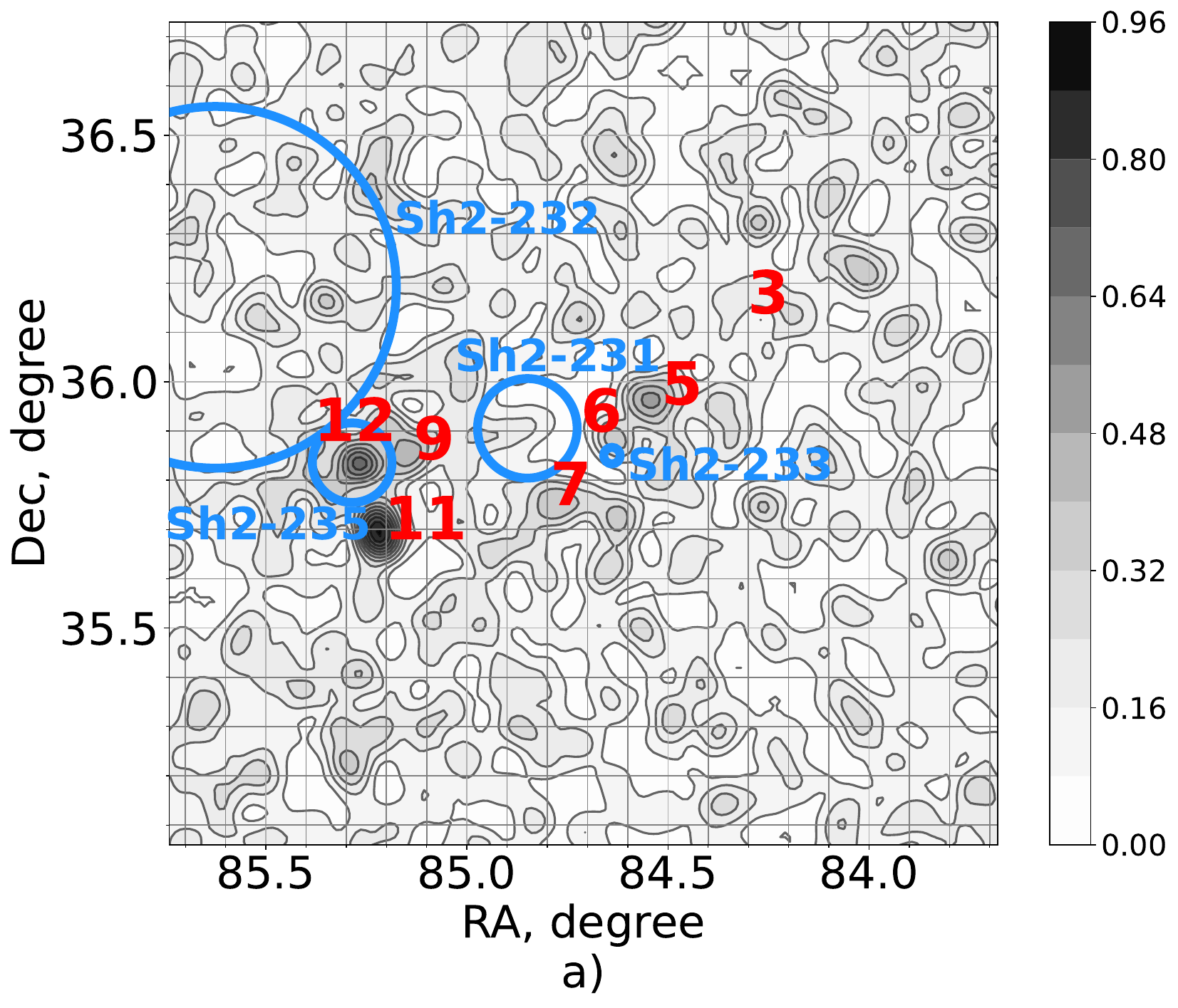}\label{FigSDMa}}
      \end{minipage}
      \hfill
      \begin{minipage}[h]{0.49\linewidth}
      \center{\includegraphics[width=1\linewidth]{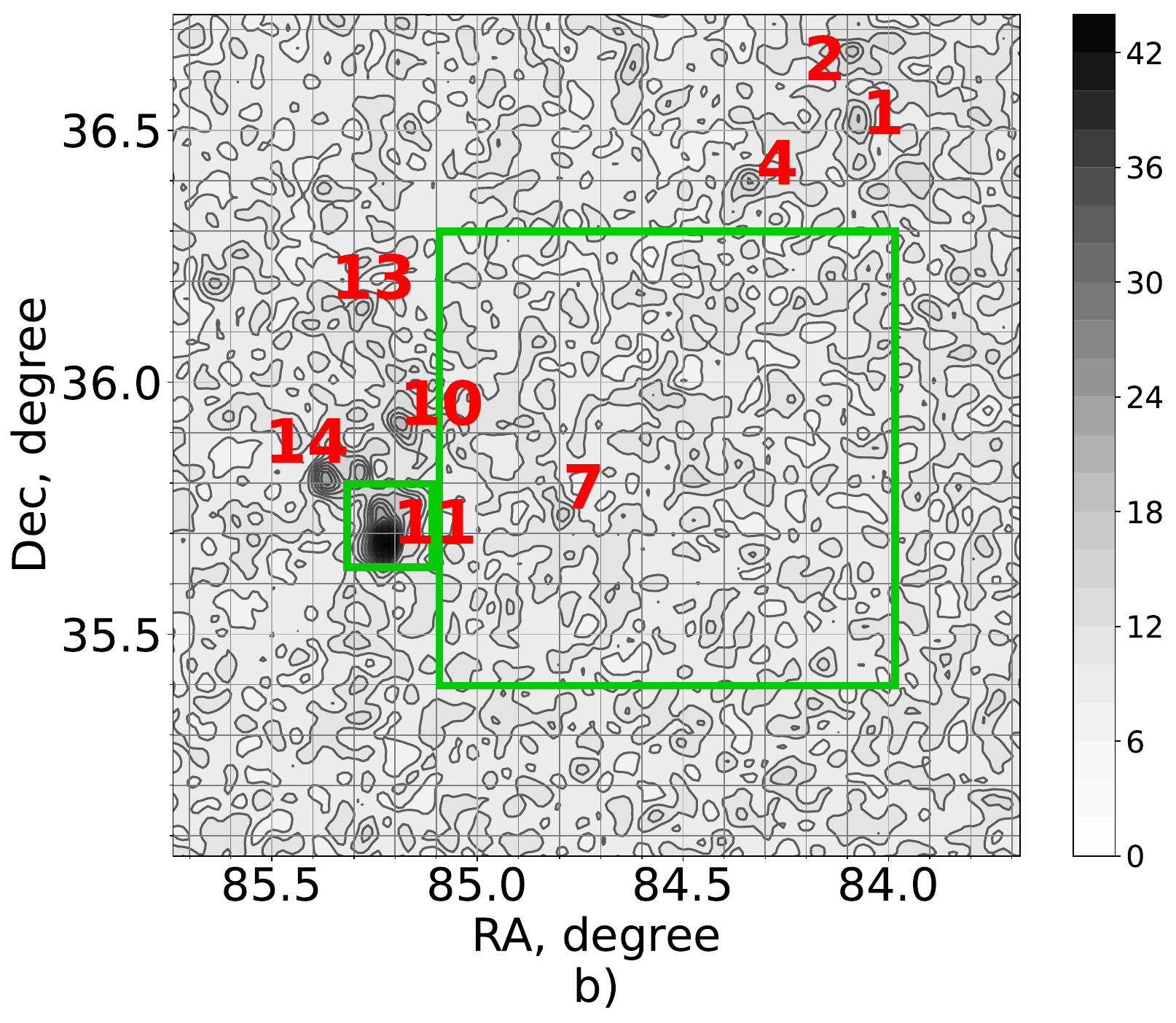}\label{FigSDMb}}
      \end{minipage}
      \vfill
      \begin{minipage}[h]{0.49\linewidth}
      \center{\includegraphics[width=1\linewidth]{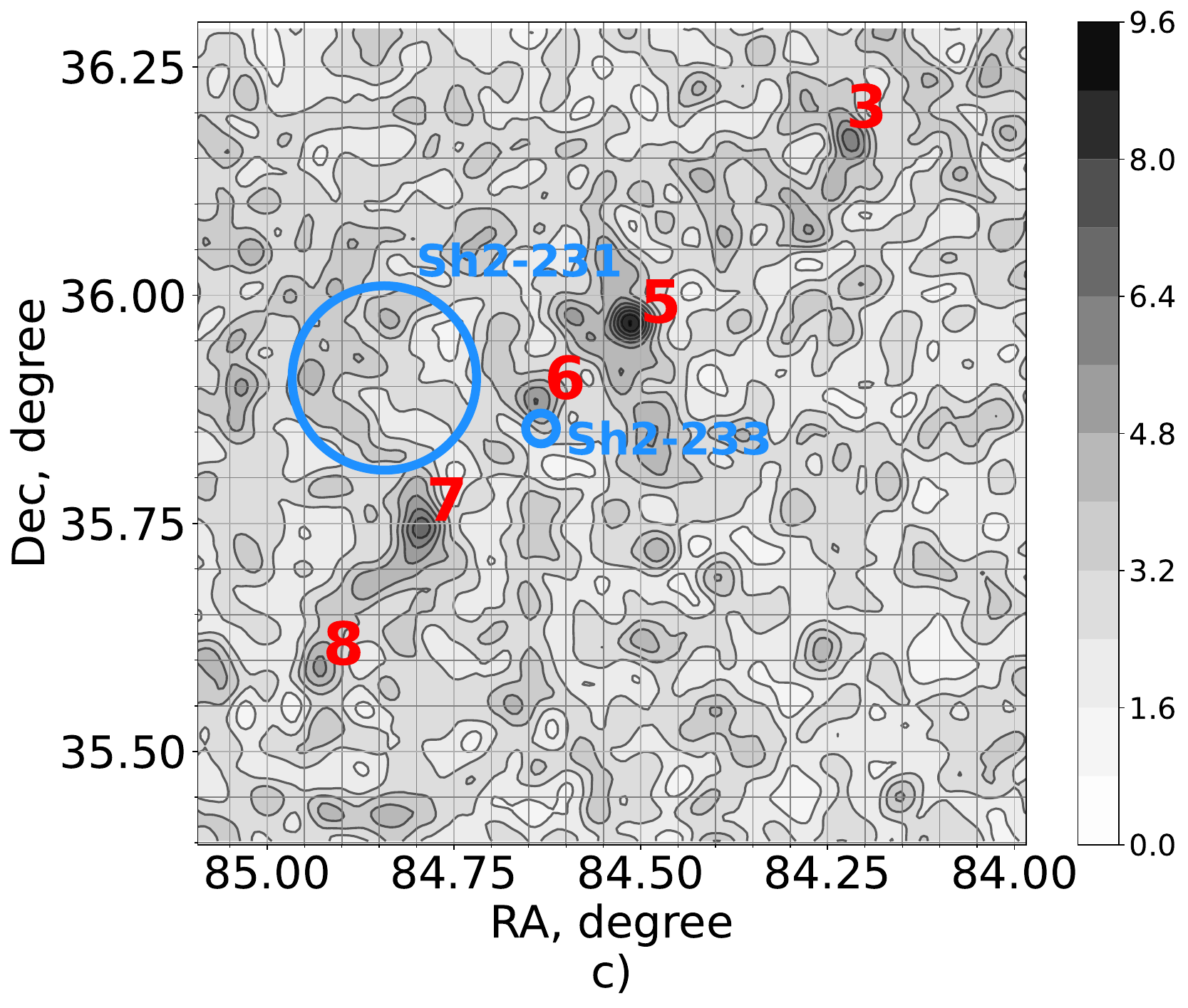}\label{FigSDMc}}
      \end{minipage}
      \hfill
      \begin{minipage}[h]{0.49\linewidth}
      \center{\includegraphics[width=1\linewidth]{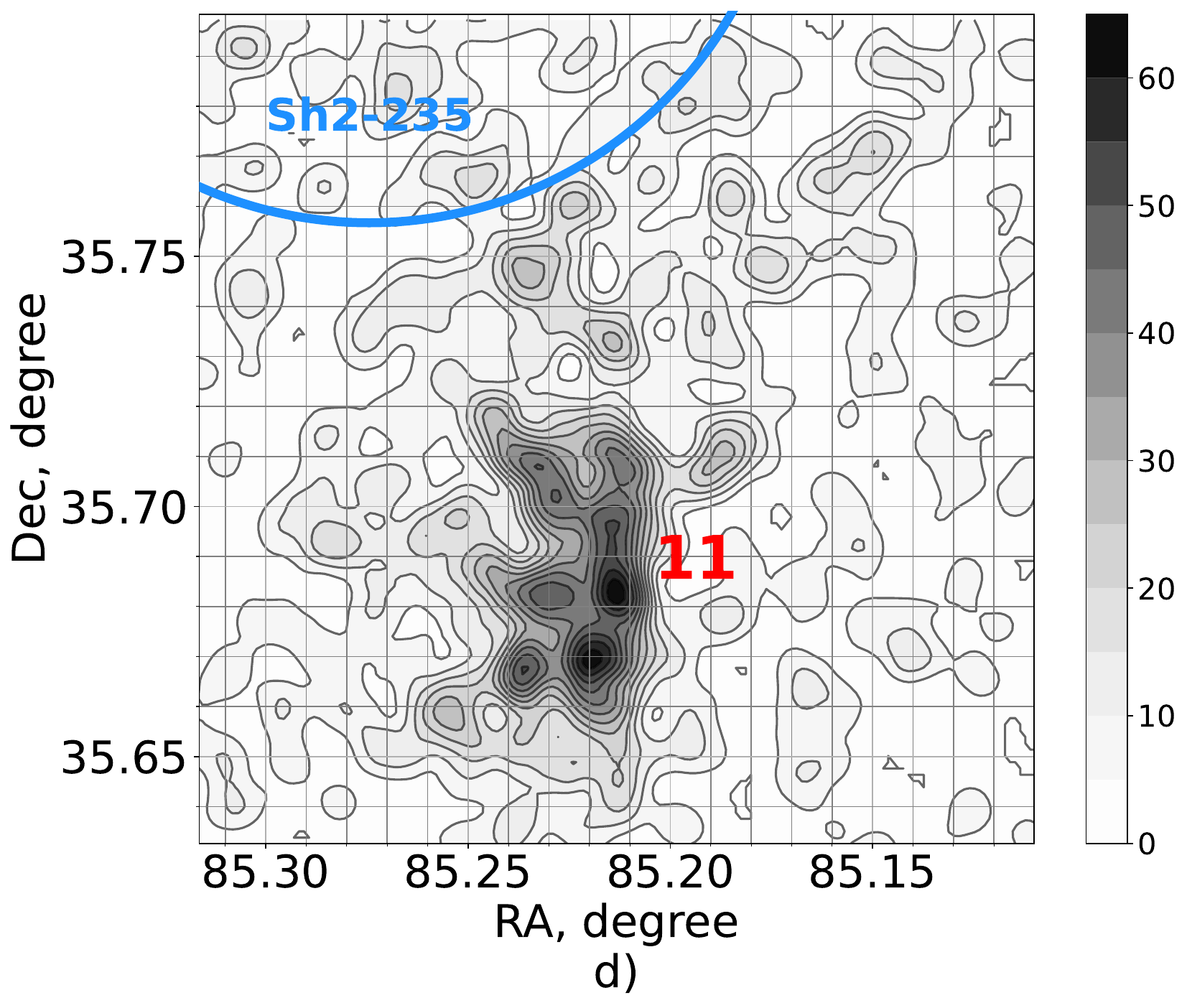}\label{FigSDMd}}
      \end{minipage}
      \caption{Stellar density maps of the G174+2.5 region based on UKIDSS data.
      Red numbers indicate areas of increased stellar density considered in the work.
      Shades of gray indicate the levels of surface density of stars (see the scales corresponding to the pictures).
      In Fig. \ref{FigSDM} (a), (c), (d), the blue circles show the HII regions, with names next to them.
      In Fig. \ref{FigSDM} (b) the green rectangles show the areas of Fig. \ref{FigSDM} (c), (d).
      a) The entire area.
      Kernel half-width $h_{KDE}=4$ arcmin, limiting magnitude $K_{0,lim}=11$.
      b) The entire area.
      Kernel half-width $h_{KDE}=2$ arcmin, limiting magnitude $K_{0,lim}=17$.
      c) The area of the WB 673 filament.
      Kernel half-width $h_{KDE}=2$ arcmin, limiting magnitude $K_{0,lim}=15$. 
      d) Area of S235~A-B-C.
      Kernel half-width $h_{KDE}=0.5$ arcmin, limiting magnitude $K_{0,lim}=15$.}
              \label{FigSDM}%
    \end{figure}

   Surface density maps were constructed using the kernel density estimator (KDE) method \citep[for a detailed description of the method, see][]{Silverman86, Seleznev16}.
   We use a biquadratic kernel with different kernel half-widths and for different limiting magnitudes $K_{0,lim}$ (Fig. \ref{FigSDM}).
   
    The stellar density maps in Fig.\ref{FigSDM} are the starting point to identify star clusters and candidate star clusters in the G174+2.5 star formation region.
    In this figure, areas of the increased stellar density identified as clusters or cluster candidates are marked with numbers.
    Some of the increased density areas are visible on both maps ($K_{0,lim} = 11$ and $K_{0,lim} = 17$), some of areas are visible only on one.
    For convenience, all the clusters we found have a common numbering in this paper.
    We summarize the correspondence between the numbers used and the names of known clusters in the Table \ref{TabClust}.
    In the same table we list the clusters and cluster candidates .
    The positions of the concentration centers on the map were determined as the coordinates of the local density maxima on the map plotted with the kernel half-width $h_{KDE} = 0.5$ arcmin and the limiting magnitude $K_{0,lim}=15$.
    
    \begin{table}[h]
    \caption{Coordinates of star clusters and cluster candidates in the star-forming region G174+2.5.}             
    \label{TabClust}      
    \centering                          
    \begin{tabular}{clcccc}       
    \hline                 
    № & Name & RA (J2000.0) & Dec (J2000.0) & R & Stars' number \\ 
    & & deg & deg & arcmin&  \\
    \hline                        
       1  &             & 84.074 & 36.522 & 2.5  &  13 \\  
       2  &             & 84.093 & 36.662 & 1.9  &  29 \\
       3  & WB89 668    & 84.221 & 36.181 & 1.4  &   9 \\
       4  &             & 84.333 & 36.391 & 2.1  &  13 \\
       5  & WB89 673    & 84.508 & 35.970 & 1.2  &   9 \\
       6  &             & 84.626 & 35.878 & 2.6  &  11 \\
       7  & S233IR      & 84.794 & 35.755 & 2.5  &  14 \\
       8  &             & 84.943 & 35.581 & 1.7  &  10 \\
       9  &             & 85.133 & 35.847 & 1.4  &  13 \\  
       10 & North-West  & 85.190 & 35.919 & 2.4  &  67 \\
          & Koposov 7   &        &        &      &     \\
          & FSR 784     &        &        &      &     \\
       11 & S235~A-B-C  & 85.214 & 35.686 & 5.5  & 326 \\
          & BDSB 71,72,73 &      &        &      &     \\
       12 & Central     & 85.278 & 35.824 & 1.3  &  35 \\
       13 & S232IR      & 85.276 & 36.152 & 0.7  &  12 \\
       14 & East1,East2 & 85.372 & 35.819 & 3.2  &  75 \\
      
    \hline                                   
    \end{tabular}
    \end{table}

     Fig.\ref{FigSDM}a and Fig.\ref{FigSDM}b show that the East2 cluster does not stand as an isolated object, but rather as a deformation  of the East1 cluster halo \textbf{(both are marked in Table \ref{TabClust} and in Fig.\ref{FigSDM}b as number 14)}.
     This result differs from what \citet{Kirsanova08} \textbf{and \citet{Camargo11}} obtained using 2MASS data.
     This can be understood if we underline that \citet{Kirsanova08} \textbf{and \citet{Camargo11}} did not account for the  effects of interstellar extinction.
     On the other hand, the map in Fig.\ref{FigSDM} cannot be  very accurate due to the statistical nature of the extinction obtained by NICEST method, which can produce density distortions on small scales.
    
    We do not see any extended stellar structure associated with the filament passing through Sh2-233 (filament WB 673 following \citet{Kirsanova17}) on the enlarged fragment of the density map of this filament region in Fig.\ref{FigSDM}c.
    Separate condensations of stars seem as being `strung' along the filament.
    However, if we can speak about the presence of some elongated structures at all, they seem to be rather oriented perpendicularly to the molecular filament and crossing it in the `knots' -- condensations of stars.
    
    Star clusters in the region of the WB 673 filament were first mentioned by \citet{Ladeyschikov16}, who showed that they lie close to the molecular gas clumps WB89 668 and WB89 673. 
    These two clumps are marked in Table \ref{TabClust} and in Fig.\ref{FigSDM}c as numbers 3 and 5, respectively.
    An extensive analysis of the stellar content of the star formation complex in GMC G174+2.5 has been done by \citet{Camargo11}, who
    explained that the S235~North-West \citep{Kirsanova08} cluster ($\#$ 10 in the table) was discovered almost simultaneously (and independently) by \citet{Koposov08} and named Koposov 7, and, somewhat earlier, by \citet{Froebrich07} in a 2MASS study, indicated as FSR 784. 
    
    A comparison of the list of star clusters given in Table \ref{TabClust} and the list of clusters from \citet{Camargo11} shows that they analyzed the 2MASS catalog with very high resolution.
    We plotted the stellar density maps using the KDE method with a kernel half-width $h_{KDE} = 4$ arcmin and $h_{KDE} = 2$ arcmin.
    Therefore, many clusters from \citet{Camargo11} fall inside the larger cluster on the maps in Fig.\ref{FigSDM} and in Table \ref{TabClust}.
    Therefore, we consider the BDSB 71, BDSB 72, and BDSB 73 clusters \citep{BDSB2003} as sub-clusters of the cluster S235~A-B-C. In particular, 
    BDSB 71 is located on the southern margin of S235~A-B-C,
    BDSB 72 falls deep inside S235~A-B-C, and 
    BDSB 73 appears as a weak condensation north to S235~A-B-C in Fig.\ref{FigSDM}d, but we consider it as a part of the S235~A-B-C halo.
    \textbf{Such differences in the identification of clusters may also be partly due to the difference in the limiting magnitude of the catalogs (UKIDSS is deeper than 2MASS) and the lack of absorption accounting by \citet{Camargo11}.
    For these reasons, we can see fainter stars than could be seen with 2MASS.
    These stars can blur and change the small-scale structure of clusters on density maps.
    This circumstance leads to a visible difference in the structure of clusters compared to previous studies based on the less deep surveys.}
    
    The complex structure of S235~A-B-C, consisting of several sub-clusters (Fig. \ref{FigSDM}d), was already noted by \citet{Kirsanova08}.
    However, the coordinates of the BDSB 71, BDSB 72, and BDSB 73 clusters do not coincide with the positions of the local density maxima in Fig.\ref{FigSDM}d, which shows the cluster region S235~A-B-C on a larger scale.
    This is not unexpected given the different spatial resolution of UKIDSS and 2MASS.
    On the other hand,  these substructures are  most probably transients, and
    dynamic evolution will eventually lead to their merging \citep{Aarseth72}.
    
    Cluster PCS 2 from \citet{Camargo11} along with S233-IR cluster falls inside the object 7 from Table \ref{TabClust}.
    On the maps, they do not differ.
    IRAS 05361+3539 is noticeable as a weak compact condensation south to object 7 on the maps in Fig. \ref{FigSDM}(a) and Fig. \ref{FigSDM}(c).
    However, on the map in Fig. \ref{FigSDM}(b) ($K_{0,lim} = 17$) it seems rather as part of the halo of cluster 7.
    The cluster CBB 1 \citep{Camargo11} on the maps in Fig. \ref{FigSDM} does not differ from IRAS 05361+3539 cluster.
    Most likely, these clusters also could be considered as sub-clusters.
    The cluster CBB 2 \citep{Camargo11} is a very faint group of stars in the nearest vicinity of S235~Central cluster (12 in Table \ref{TabClust}), northeast of it.
    We do not identify CBB 2 as an independent object in this work.
    
\subsection{Determining cluster parameters}\label{4.2}

    We study the structure of the clusters using the radial density profiles plotted by the KDE method with a kernel half-width of 0.5 arcmin and a limiting magnitude $K_{lim}=15$.
    Radial density profiles of all clusters in the G174+2.5 region exhibit a maximum, significantly larger than the background level.
    
\begin{figure}
      \centering
      \includegraphics[width=17cm]{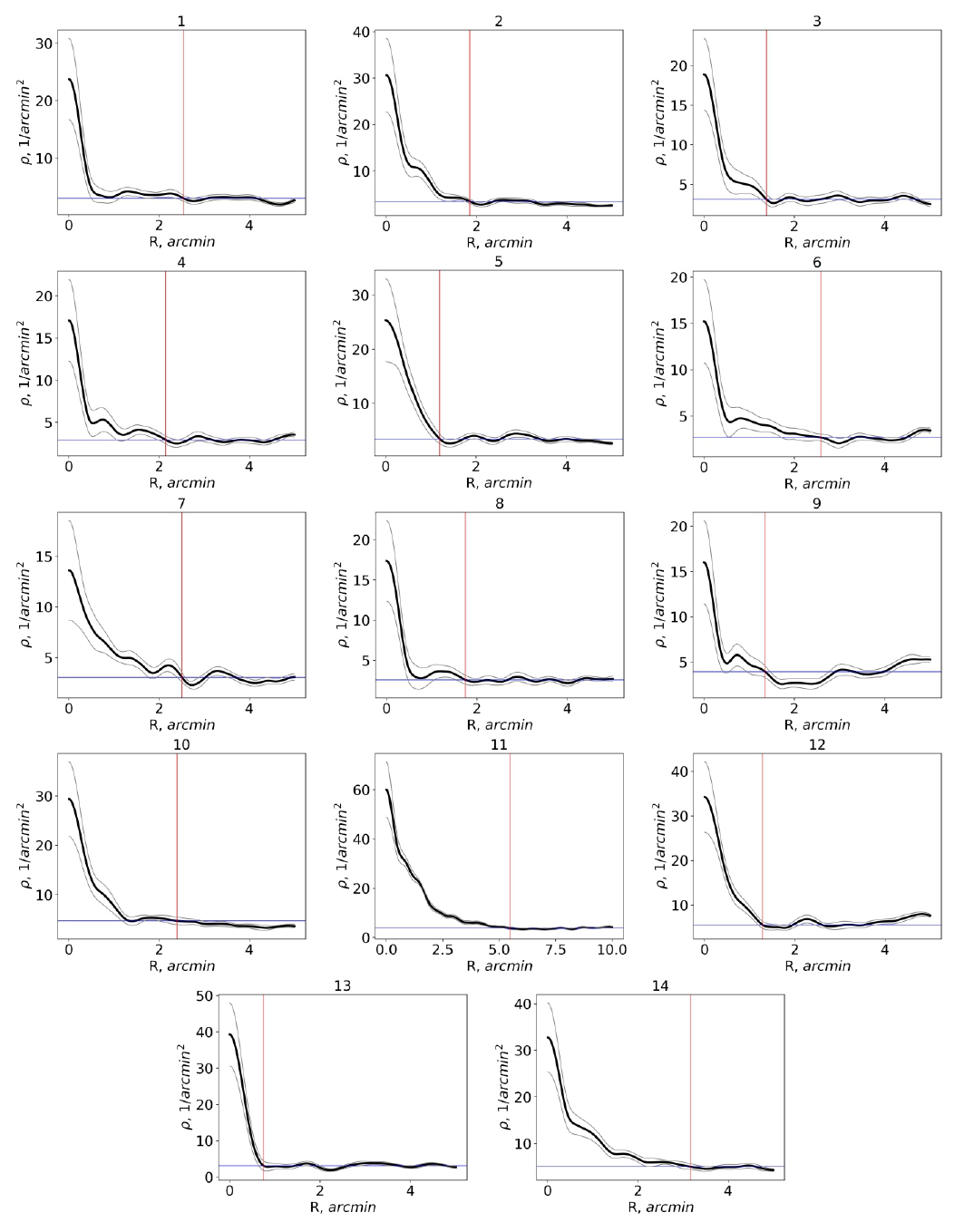}      
      \caption{Radial density profiles of the clusters.
      The profile is shown by thick black line, the confidence interval is shown by thin gray lines, the cluster radius position is marked by red, and the background level is plotted by blue.
      The number of cluster is shown at the top of each panel.}
              \label{FigClustProf}%
    \end{figure}

    The cluster radius is determined from the radial density profile under the assumption that the cluster is spherically symmetric.
    The profile is built for distances from the cluster center that are several times larger than the apparent size of the cluster.
    Next, we consider the part of the density profile which obviously belongs to the background. 
    The background level is drawn so that the areas of the density profile fluctuations above and below this level are the same.
    Then the cluster radius is defined as the abscissa of the first intersection of the background level with the radial density profile.
    To obtain the confidence interval for the radial density profile and determine the radius error, secondary samples (20 pieces) are formed with the same density distribution as the original one.
    Next, using the KDE method, new radial density profiles are constructed from these secondary samples.
    The spread of density values at nodes characterizes the confidence interval of the original profile.
    The cluster radius error was determined as the abscissa of the intersection of the background level with the lower confidence interval of the profile.
    This method is described in more detail in \citet{Seleznev16}, and
    the obtained radii are shown in Table \ref{TabClust}, while
    the radial density profiles of the clusters are shown in the Fig.\ref{FigClustProf}.

    \begin{figure}
      \centering
      \includegraphics[width=17cm]{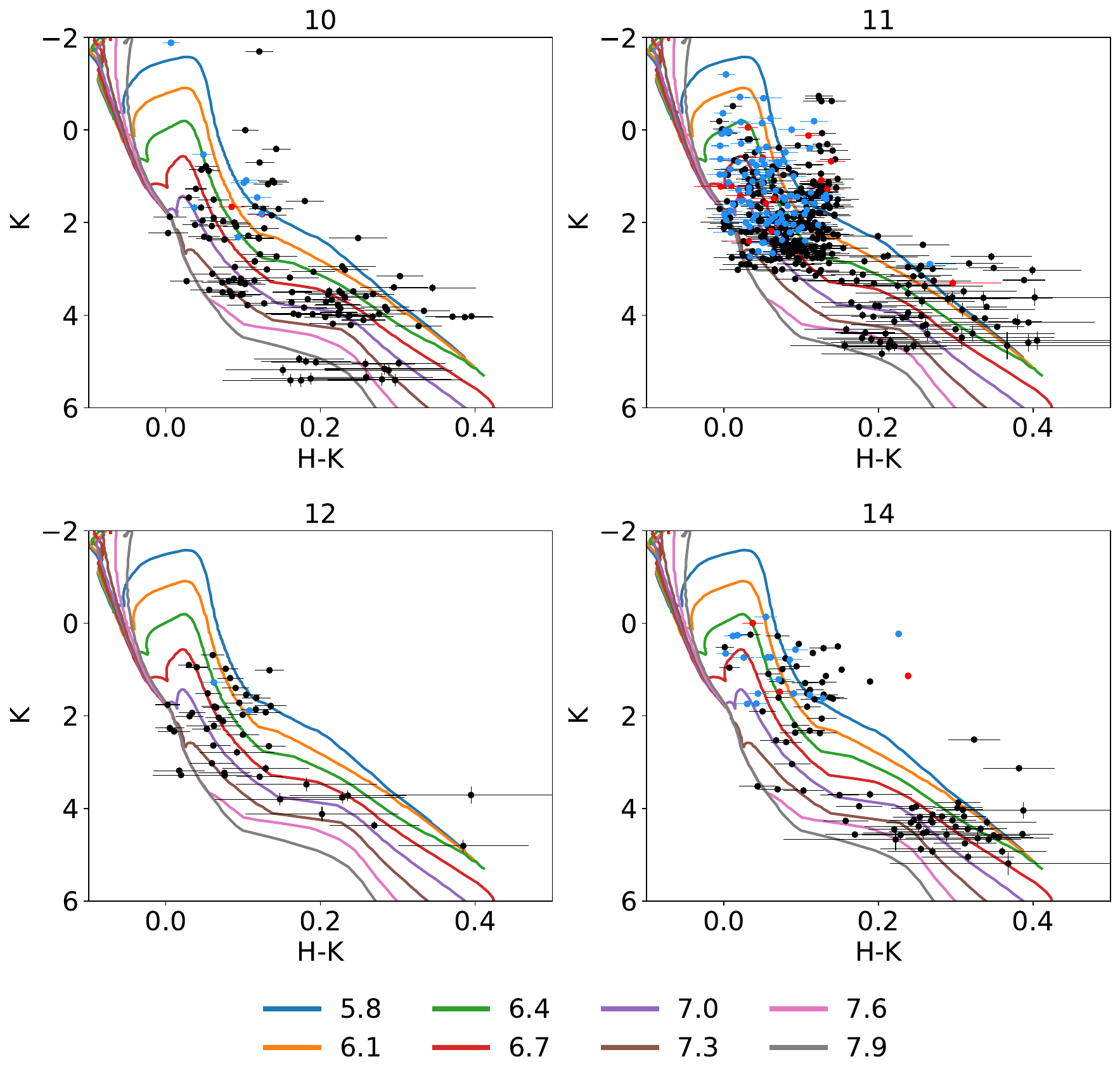}
      \caption{Color - magnitude diagrams of clusters 10 (S235~North-West), 11 (S235~A-B-C), 12 (S235~Central) and 14 (S235~East1+East2), combined with isochrones \citep{Bressan12}.
      The color excesses and absorption in the K band were obtained by the Q-method (see description in Appendix \ref{Ap}).
      Class I YSOs are shown in red, class II YSOs in blue, other stars in black.
      Colored lines show the isochrones, different colors indicate different decimal logarithms of age (see a legend).}
              \label{FigCMD_Q}
    \end{figure}

    \begin{figure}
      \centering
      \centering
      \includegraphics[width=17cm]{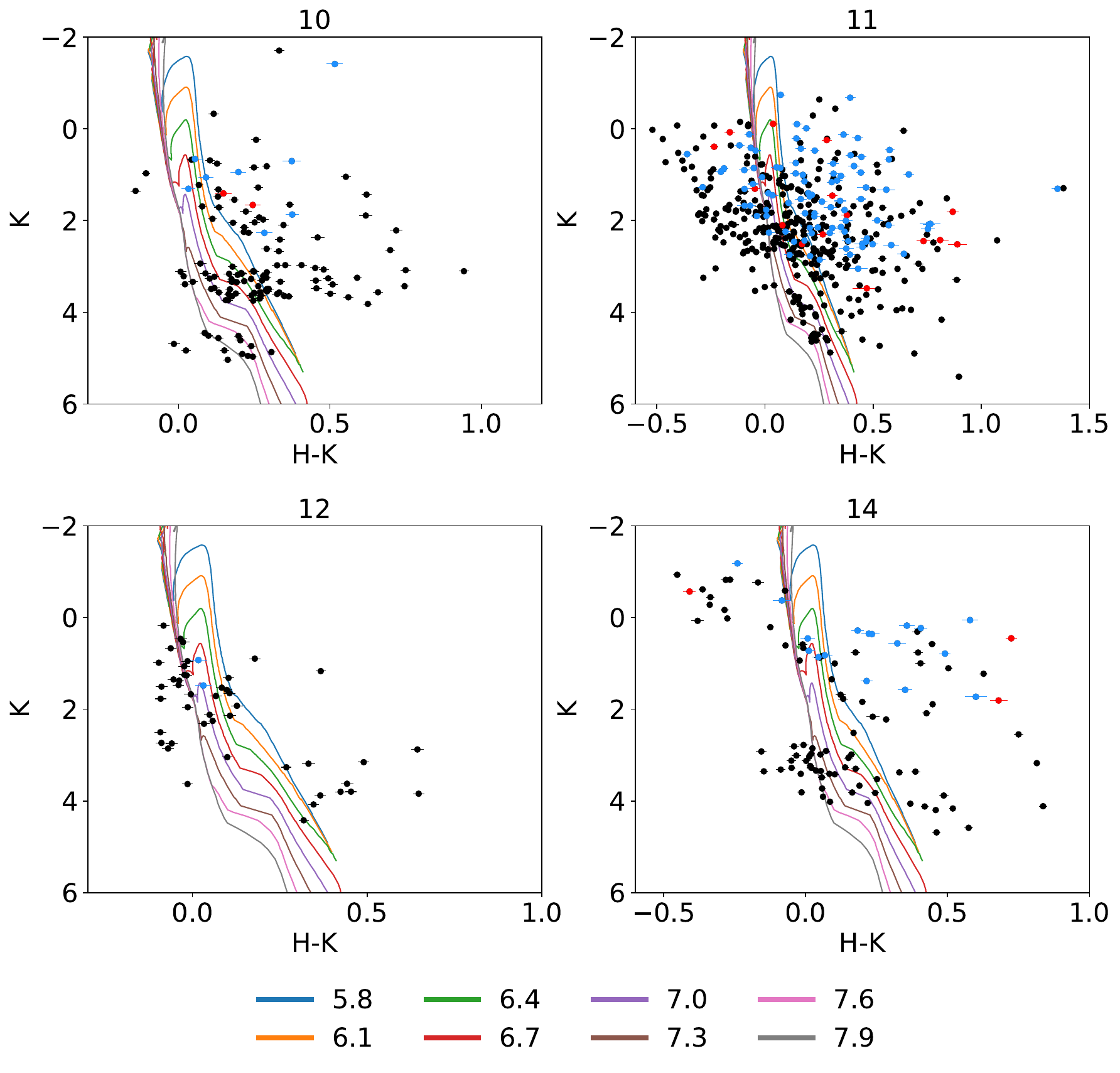}
      \caption{Color - magnitude diagrams of clusters 10 (S235~North-West), 11 (S235~A-B-C), 12 (S235~Central) and 14 (S235~East1+East2), combined with isochrones \citep{Bressan12}.
      The color excesses and absorption in the K band were obtained by the NICEST method (see Section \ref{3}).
      Designations are the same as in Fig.\ref{FigCMD_Q}.}
              \label{FigCMD_N}%
    \end{figure}
    
    For the cluster areas, limited by the circles with the previously determined radii (see Fig. \ref{FigExtMap}), the number of stars in the cluster was estimated by comparison with the field regions where there are no cluster stars.
    That is, we subtract the number of stars in the region that includes only the field stars from the number of stars in the region with the cluster.
    For some areas of clusters, the comparison region is a ring around the cluster circle, the area of a ring is equal to the area of the circle.
    For the clusters in close groups, we select several comparison areas outside the cluster group.
    Every comparison area is a circle equal in area to the cluster circle.
    Then we average the number of stars in the comparison fields.
    We consider only stars with $K<15$ to determine the cluster star number, since the surface density maps for this limiting magnitude show the largest contrast for most clusters.
    Otherwise (with a larger limiting magnitude), this method can give negative values for the number of stars in some cases.
    The number of stars in clusters with the membership probabilities (see below) was determined by a direct counting.
    We consider the threshold membership probability of a star of 0.7 to be counted as a cluster member.
    The results of the star counts are shown in Table \ref{TabClust}.
    
    We use the UPMASK package \citep{Krone14} to estimate the membership probability of stars in the clusters.
    UPMASK does not use any model hypothesis.
    The only assumption is the spatial homogeneity of the field star density in the considered area comparing to the cluster.
    Although the coordinate space plays an important role in the UPMASK method, photometric parameters (or information about proper motions and parallaxes) are also used.
    Initial clustering is performed using the information contained in the photometry.
    After this, the clustering of the selected groups in the coordinate space is checked.
    Thus, for a real field with some density fluctuations, in the photometric space the field stars will not be related, since they do not have a common origin.
    Therefore, they will not be selected as a cluster at the beginning in most iterations of the algorithm and will therefore receive a low membership probability.

    For our clusters we used the following parameters of the UPMASK algorithm.
    The minimum number of stars per cluster was chosen to be 20 in all cases.
    \citet{Krone14} obtained the best cluster detection results with values between 10 and 25 stars per clustering.
    In our case, when we use smaller values in clusters 10 (S235~North-West) and 12 (S235~Central), UPMASK cannot estimate the membership probabilities of the stars.
    Therefore, we adopt a larger value for the parameter of stars per cluster.
    To obtain the probability estimate, 2,000 runs of the algorithm are performed for stars with known photometry in all three UKIDSS bands (J, H, K).
    We use the equatorial coordinates of stars, magnitudes in the J, H and K bands, color indices J-H, J-K, and H-K and the errors in these magnitudes as the input data for UPMASK.
    
    This algorithm is used only for clusters with well-defined clustering of stars on the maps, namely 10 (S235~North-West), 11 (S235~A-B-C), 12 (S235~Central) and 14 (S235~East1+East2).
    Only information about stars located in regions containing clusters is used to estimate membership probabilities.
    For clusters 11 (S235~A-B-C) and 14 (S235~East1+East2), such regions are circles with centers coinciding with the centers of the clusters and radii equal to the radii of the clusters.
    For clusters 10 (S235~North-West) and 12 (S235~Central) we had to use circles with a radius of twice the cluster radius, since the smaller regions did not contain enough stars for UPMASK to work.
    We cannot estimate the membership probability for clusters 5 (WB89~673) and 6, due to insufficient photometric data for stars in this region (see Fig.\ref{FigHJmis}).
    We cannot process the cluster areas 1, 2, 3 (WB89~668), 4, 7 (S233IR), 8, 9, and 13 (S232IR) with UPMASK either, because of low number statistics.

    Here and below we assume that clusters 10 (S235~North-West), 11 (S235~A-B-C), 12 (S235~Central) and 14 (S235~East1+East2) include stars with membership probabilities greater than 0.7.
    There is currently no generally accepted agreement on what membership probability threshold is best to use.
    An estimate of 0.7 is quite strict and allows us to exclude most background stars, which can significantly affect the estimate of cluster parameters.
    
    For clusters 10 (S235~North-West), 11 (S235~A-B-C), 12 (S235~Central) and 14 (S235~East1+East2), an attempt was made to estimate the distance modulus.
    To do this, we superimposed theoretical isochrones to the absorption-corrected photometric sequences of clusters (Fig.\ref{FigCMD_Q} and Fig.\ref{FigCMD_N}).
    We use the solar metallicity isochrones for UKIDSS filters in the J, H, and K bands from the Padova suite of models \citep{Bressan12}.
    The combination of sequences with isochrones was carried out in such a way that the majority of stars were between isochrones with a decimal logarithm of age 6.1 and 6.7.
    
    To correct magnitudes for extinction, data on the color excesses of stars in clusters were obtained by two methods: (1) the Q-method (see Appendix \ref{Ap}) and (2) by the total extinction in the K band, estimated with the NICEST method (see section \ref{3}).
    To do this we use the coefficients for the total extinctions $A_J/A_V=0.28170$, $A_H/A_V=0.18251$, and $A_K/A_V=0.11281$ listed in the output page of an isochrone\footnote{\url{http://stev.oapd.inaf.it/cmd}} \citep{Bressan12,Cardelli89,O'Donnell94}.
    It is worth noting here that due to the uneven coverage of the UKIDSS catalog in the J and H bands, it is impossible to determine the absorption by the Q-method for some stars and they fall out of our sample.
    However, the absorption maps in cluster regions plotted using the Q-method data show a similar reddening distribution to the maps plotted using the NICEST data for UKIDSS photometry (see Section \ref{3}).
    The results of the isochrone fitting for the absorption-corrected cluster sequences are shown in Fig.\ref{FigCMD_Q} (for the Q-method) and Fig.\ref{FigCMD_N} (for the NICEST method).
    
    Since the clusters are extremely young, they contain a large number of the pre-main-sequence stars.
    Consequently, the clusters have fairly wide photometric sequences along the magnitude axis, which makes it difficult to match them with isochrones.
    We fit the isochrones to the cluster sequence by eye so that the points of the sequence lie mostly above the isochrone for $\log (t) = 6.7$.
    In Fig.\ref{FigCMD_N}, the cluster sequences have a considerable number of stars far from the isochrones along the color index axis, and the sequences themselves are rather diffuse.
    The large scatter of points in the color-magnitude diagrams (CMDs), when we use the extinctions obtained by NICEST, is due to the fact that this method does not take into account the deviations of the reddening values of individual stars from the average in the region.
    For this reason, we cannot use such diagrams to determine distances to clusters.
    When we use the individual reddening values estimated with the Q-method, the scatter on the CMDs becomes significantly smaller  (Fig.\ref{FigCMD_Q}).
    In this case the cluster sequences are located within the range of color indices of the isochrones and it is possible to estimate the distance to the clusters (see Table \ref{TabClustPar}).
    To obtain the error, we shift the photometric sequences of clusters by the color index error values. 
    Next, we determine the distance to the cluster using these shifted sequences. 
    We take the difference between these values and the original distance as an error.

    Another method to determine the cluster distance is the use of stellar parallaxes from the Gaia DR3 catalog.
    To estimate the distance, we use the inverse of the average parallax of cluster member stars.
    The results are listed in Table \ref{TabClustPar}.
    We list there the distance estimate obtained by radio-interferometry for the maser source in the cluster S235~A-B-C \citep{Burns15} as a reference point for comparison.
    Despite the photometric method is approximate, our distance estimates are consistent with the previously obtained values \citep{Foster15, Evans81}.
    Our photometric distances have larger scatter than distances from parallaxes.
    It is important, that our values obtained from parallaxes agree well with \citet{Burns15} value obtained from the maser sources trigonometric parallaxes.
    
    \begin{table}[h]
    \caption{Extinction, distances (in kpc) and proper motions of the clusters}
    \label{TabClustPar}      
    \centering                          
    \begin{tabular}{cccccccc}       
    \hline
    № & \multicolumn{2}{c}{$A_k$} & Photometric distance & Distance from Gaia $\varpi$ & Distance \footnote{Distance to the $H_2O$ masers in S235AB-MIR \citep{Burns15}} & pmRA & pmDec\\ 
     & (Q-method)& (NICEST) & kpc & kpc & kpc & mas/yr & mas/yr \\ 
    \hline                        
    10 & $0.92\pm0.07$ & $0.76\pm0.02$ & $1.7\pm0.3$ & $1.9\pm0.4$ &  & $0.50\pm0.36$ & $-2.59\pm0.31$ \\  
    11 & $1.25\pm0.05$ & $1.13\pm0.02$ & $1.9\pm0.3$ & $1.6\pm0.2$ & 1.56 & $0.06\pm0.18$ & $-2.82\pm0.15$  \\
    12 & $0.97\pm0.09$ & $0.97\pm0.03$ & $2.2\pm0.4$ & $1.7\pm0.4$ &  & $0.35\pm0.23$ & $-2.80\pm0.25$  \\
    14 & $0.82\pm0.11$ & $0.88\pm0.02$ & $1.7\pm0.2$ & $1.6\pm0.2$ &  & $0.26\pm0.26$ & $-2.82\pm0.22$  \\
    \hline                                   
    \end{tabular}
    \end{table}

\section{Kinematics}\label{5}

    We take advantage of Gaia DR3 data to investigate the clusters' kinematics.
    The mean values of the proper motions of the clusters are listed in Table \ref{TabClustPar}.

\begin{figure}
      \centering
      \includegraphics[width=17cm]{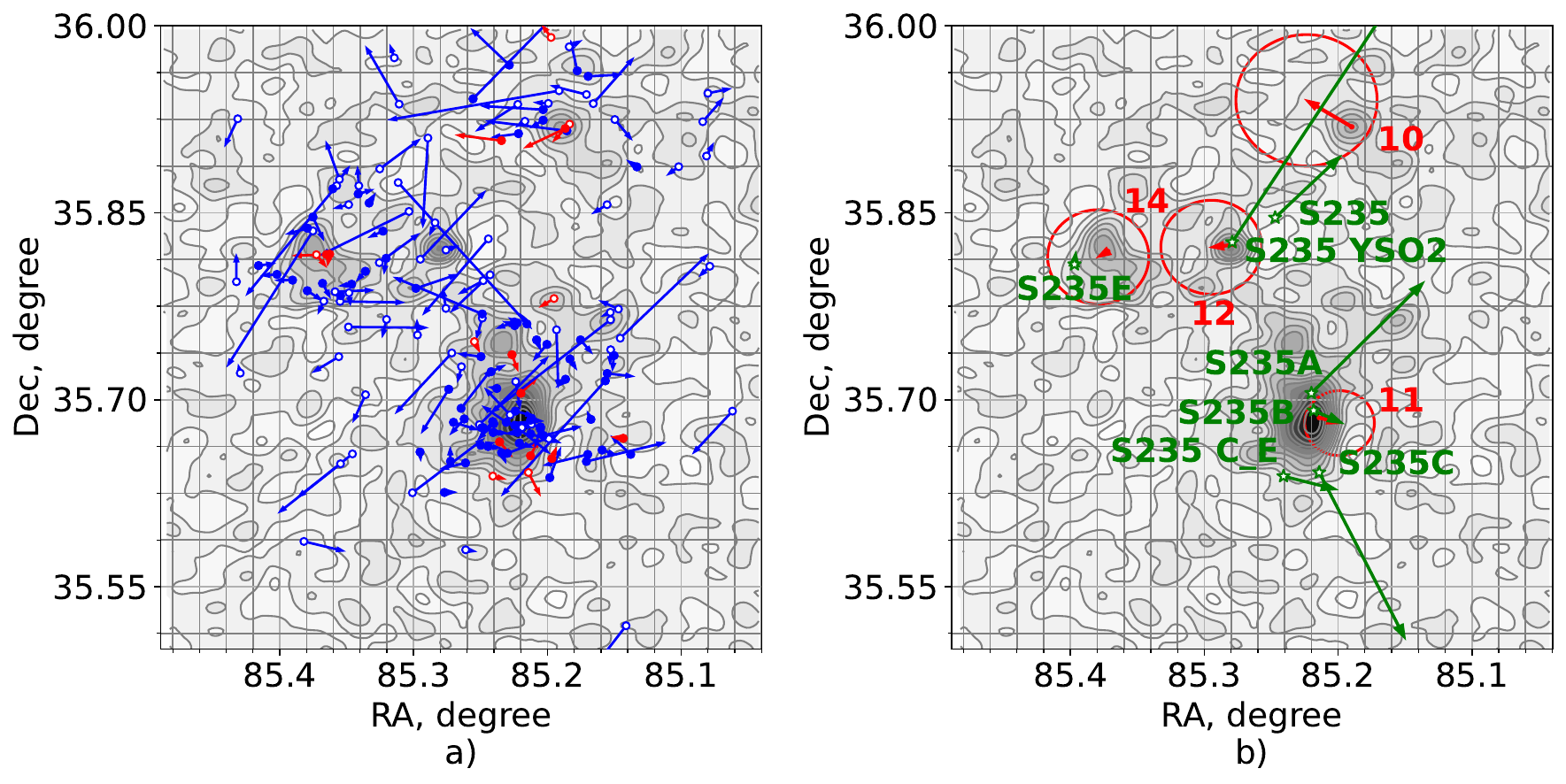}
      \caption{The left figure shows a map of residual velocities of YSO.
      Class I YSOs are shown in red, class II YSOs in blue.
      Filled markers are YSO of the clusters, unfilled markers are YSO of the field.
      The right figure shows a map of the residual velocities of clusters (red) and massive stars (green) in the region. 
      The circles show the errors.
      Velocities are given in degrees/year (for display in the figure, values are multiplied by 40000 for YSO and by 400000 for clusters).}
              \label{FigVelYSO}%
    \end{figure}

    To study the kinematics of young stellar objects (YSO) in clusters, we plot their residual velocity (Fig. \ref{FigVelYSO}a).
    The residual velocity is calculated as the difference between the YSO velocity and the average velocity of all stars in clusters 10 (S235~North-West), 11 (S235~A-B-C), 12 (S235~Central) and 14 (S235~East1+East2) with membership probabilities greater than 0.7.
    In general, the tangential velocities of young stellar objects are distributed chaotically.
    However, in the regions of clusters 10 (S235~North-West), 11 (S235~A-B-C) and 14 (S235~East1+East2) there are a small number of YSOs with velocity vectors directed away from the centers of the corresponding clusters.
    This may indicate a possible partial expansion of the young star subsystem.
    Further study of the kinematics using plots `velocity components~-- position of star' did not reveal any signs of cluster expansion (see below). 
    Nevertheless, we found in cluster 14 (S235~East1+East2) a deviation from the motion of the cluster as a whole.

    Additionally, we considered the residual velocities of clusters with respect to the same centroid.
    In Fig. \ref{FigVelYSO}b it is seen that the velocities of clusters 12 (S235~Central) and 14 (S235~East1+East2), located on the boundary of the HII region, are quite close and directed away from the center of this region.
    The residual velocity vector of cluster 10 (S235~North-West) is directed tangentially to the boundary of the HII region.
    In cluster 11 (S235~A-B-C), the velocity vector direction is opposite to other clusters in the region.
    Most likely, this can be associated with the fact that cluster 11 (S235~A-B-C) formed independently of other clusters in the region, whose formation is possibly associated with the expansion of the HII shell \citep{Kirsanova08,Kirsanova14}.
    Unfortunately, the errors in the residual velocities of the clusters are large (especially for clusters 12 (S235~Central) and 14 (S235~East1+East2)), which does not allow us to talk about the statistical significance of this result.

    \begin{table}[h]
    \caption{Massive stars in the region of Sh2-235}
    \label{TabMassStar}      
    \centering                          
    \begin{tabular}{lcccccccc}       
    \hline
    Star   & \multicolumn{3}{c}{RA} & \multicolumn{3}{c}{Dec} & Other name    & Ref \\ 
           &   h    &   m   &   s   & $^\circ$ &   '   &   ''   &               &   \\ 
    \hline                        
    S235   &  05    &  40   &  59.3 &  +35   &   50  &   46   &BD+35$^\circ$1201&[1]\\  
    S235A  &  05    &  40   &  52.8 &  +35   &   42  &   19   &               & [2] \\
    S235B  &  05    &  40   &  52.4 &  +35   &   41  &   29   &               & [2] \\
    S235C  &  05    &  40   &  51.4 &  +35   &   38  &   30   &               & [3] \\
    S235C\_E& 05    &  40   &  57.8 &  +35   &   38  &   19   &               &     \\
    S235E  &  05    &  41   &  35.3 &  +35   &   48  &   31   &               &     \\
    S235 YSO1 & 05  &  41   &  11.1 &  +35   &   50  &   02   &  IRS2         & [1] \\
    S235 YSO2 & 05  &  41   &  07.0 &  +35   &   49  &   35   &  IRS1         & [1] \\
    \hline
    \multicolumn{9}{c} {\it Notes} [1] \citet{Dewangan16}, [2] \citet{Dewangan11},   [3] \citet{Chavarria14} \\
    \end{tabular}
    \end{table} 

    Fig. \ref{FigVelYSO}b also shows the residual velocities of the massive stars in the region (see Table \ref{TabMassStar}).
    These stars are really massive because zones of ionized hydrogen form nearly all of them.
    It is well seen in the images in R band and in 3.6 $\mu$m.
    S235B also is a massive star, it was shown in \citet{Boley+2009}.
    Stars S235~YSO1 and S235~YSO2 (from here on the numbers 1 and 2 are part of the designation of these massive stars, classes of young stellar objects will be indicated by numbers I and II) are very bright in the IR, consequently, they have large luminosity and large mass.
    
    We obtained velocities of these stars by a cross-correlation with Gaia DR3 using TOPCAT (data on star S235~YSO1 are absent in Gaia DR3).
    The correspondence for star S235 was found at the distance of about 2 arcseconds (the star of 10 mag in G band), for star S235E~--- at the distance of about 1.5 arcseconds (the star of 13 mag in G band), for the rest of the stars~--- at the distance of about 1 arcsecond (the stars of 16-19 mag in G band).
    
    The comparison with the list of young stellar objects and the list of the cluster members shows that star S235A belongs to cluster S235~A-B-C (11), it is young stellar object of class I.
    Stars S235C and S235C\_E also are young stellar objects of class I but are not cluster members according to our estimates.
    Other stars are not in the available list of YSOs \citep{Chavarria14}.
    However, when we compared the positions of massive stars with the list of YSOs from \citet{Dewangan11}, we found that S235B most likely also belongs to class I.
    For the rest of the stars we did not find information about their classes.
    Star S235 is outside the cluster regions, stars S235B and S235E are not cluster members, stars S235~YSO1 and S235~YSO2 both are members of cluster S235~Central (12).
    The membership (non-membership) of these stars in clusters by our estimates is rather questionable because many of them are surrounded by bright nebulosities, and their magnitudes and colors could be distorted.
    
    Stars S235B and S235C\_E have the small velocities with respect to cluster S235~A-B-C.
    Star S235E have a small velocity with respect to the cluster S235~East1+East2.
    Star S235~YSO2 have a large velocity with respect to the cluster S235~Central.
    The large velocities of stars S235A and S235C with respect to cluster S235~A-B-C may be an argument in favor of the fact that S235~A-B-C could not be gravitationally bound (see below).
    
    To study the clusters' internal kinematics, we consider the radial and tangential velocity components of the cluster members.
    These velocity components are plotted relative to the center of the cluster in the tangential plane.
    The dependence of these velocity components on the position of stars in the cluster can help to track possible rotation of the cluster, its compression or expansion, possible irregularities of motion within the cluster, and differences in the velocity distribution of YSOs and other stars in the cluster. Here, we mean the position of the star in the cluster as the distance from the center of the cluster and the position angle of the star relative to the $RA$ axis.
    
    In the case of the translational motion of the cluster stars in the absence of any internal motions of the stars, we should see a clear sinusoidal sequence on the plots `velocity component~-- positional angle'. 
    This sinusoidal sequence should have the same amplitudes of maximum and minimum and the distance of $\pi$ radians between the points of intersection of the sequence with the abscissa axis.
    When we add some bulk rotation over the translational motion, the sinusoidal sequence on the plot `tangential component~-- positional angle' will show a displacement up or down along the ordinate axis (the tangential velocity component). 
    When we include expansion or contraction of the cluster kinematics, we will see a similar displacement in the `radial component~-- positional angle' plot.
    If the apex of the cluster stars motion is located close to the cluster (projected onto the celestial sphere), or the cluster has some other internal motion, the humps of the sinusoidal sequence becomes asymmetrical.
    When the cluster stars in the outer parts of the cluster have a larger velocity dispersion comparing to the cluster core, the plots `velocity components~-- distance' will show an increase in the scatter of values with distance.
    
\begin{figure}
      \centering
      \includegraphics[width=17cm]{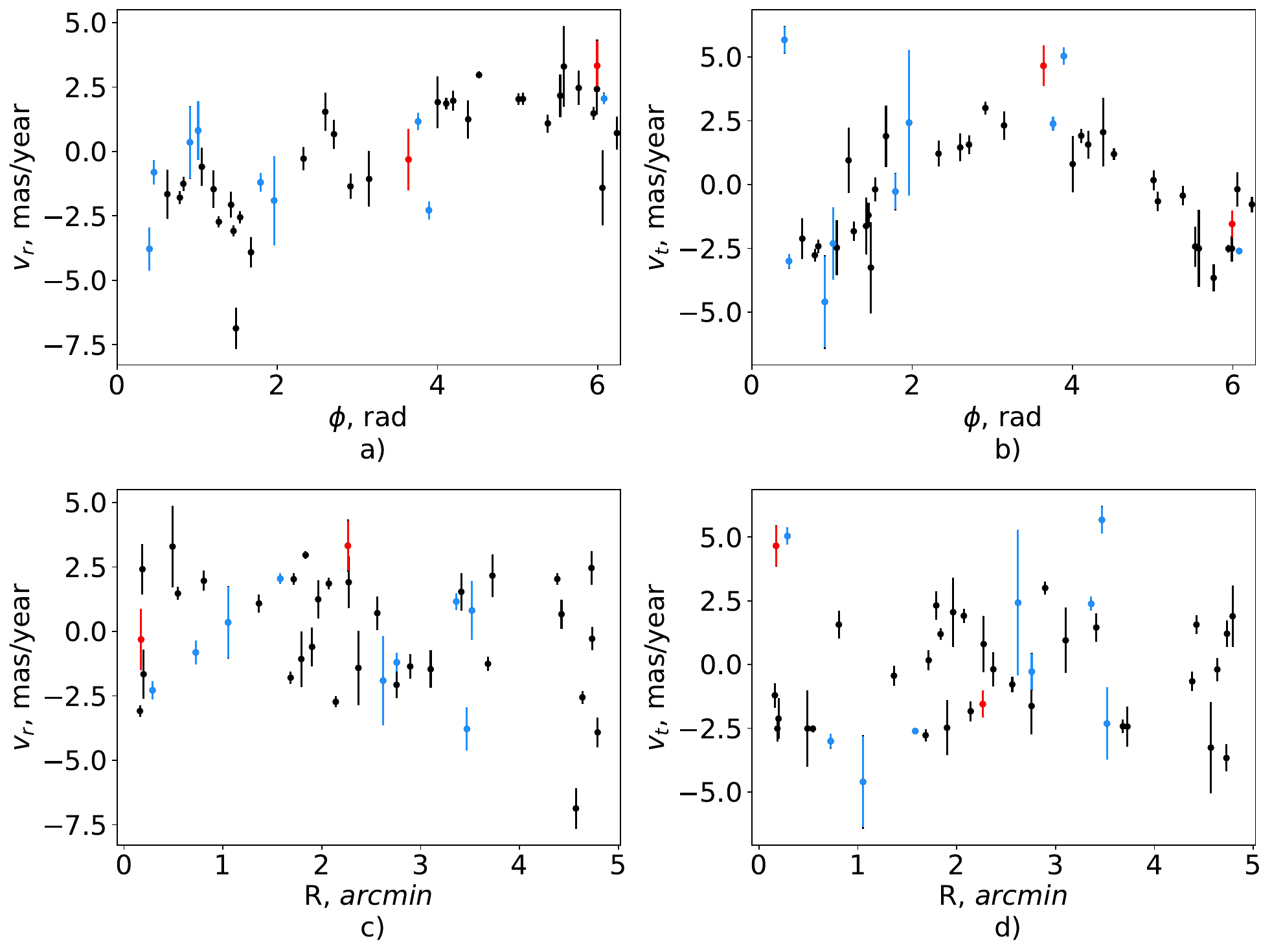}
      \caption{Plots `tangential component -- positional angle', `radial component -- positional angle', `tangential component -- distance' and `radial component -- distance' for cluster 10 \textbf{(S235~North-West)}. 
      Class I YSOs are shown in red, class II YSOs in blue, other stars in black.}
              \label{FigVC10}%
    \end{figure}

\begin{figure}
      \centering
      \includegraphics[width=17cm]{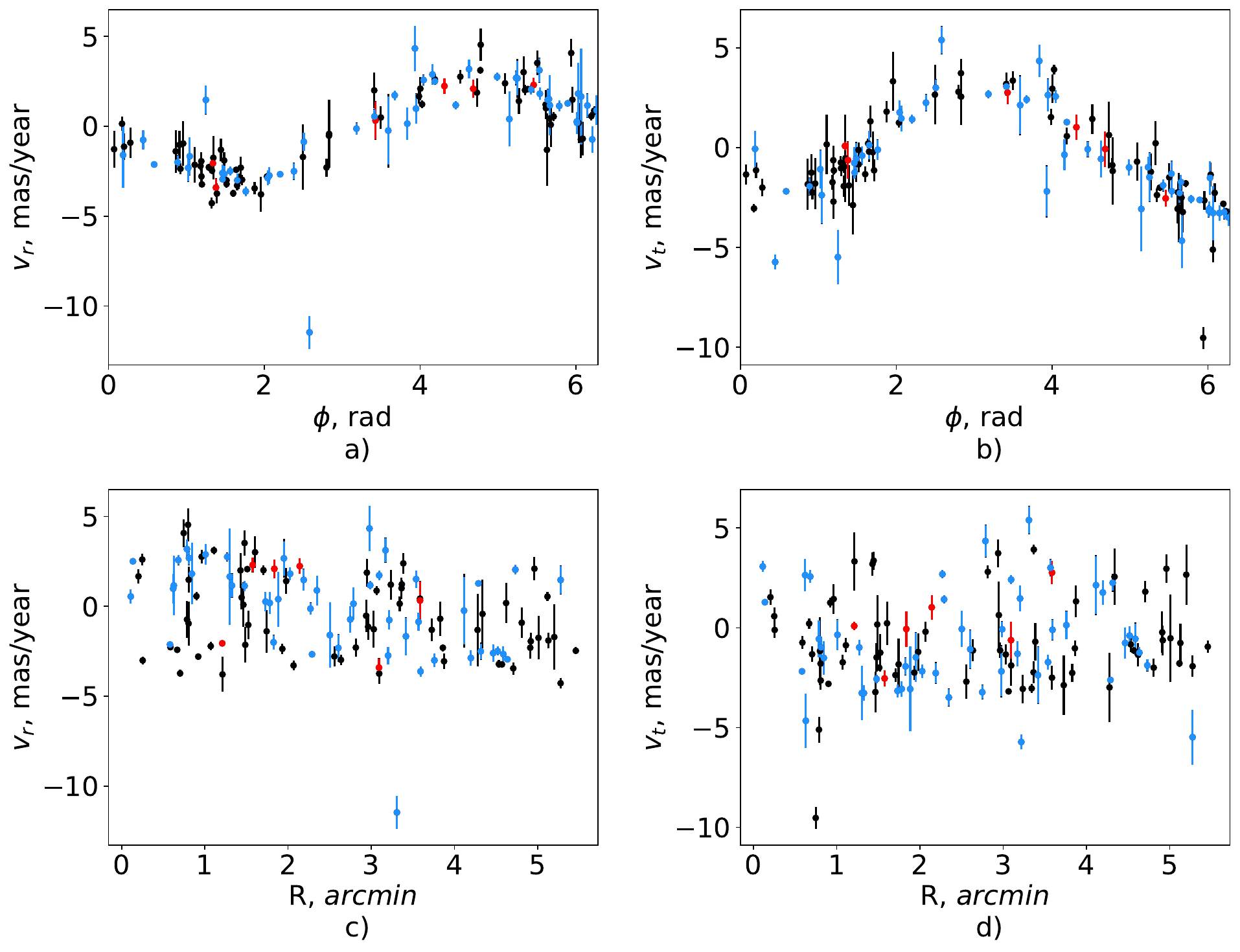}
      \caption{The same as Fig.\ref{FigVC10} for cluster 11 (S235~A-B-C).}
              \label{FigVC11}%
    \end{figure}

\begin{figure}
      \centering
      \includegraphics[width=17cm]{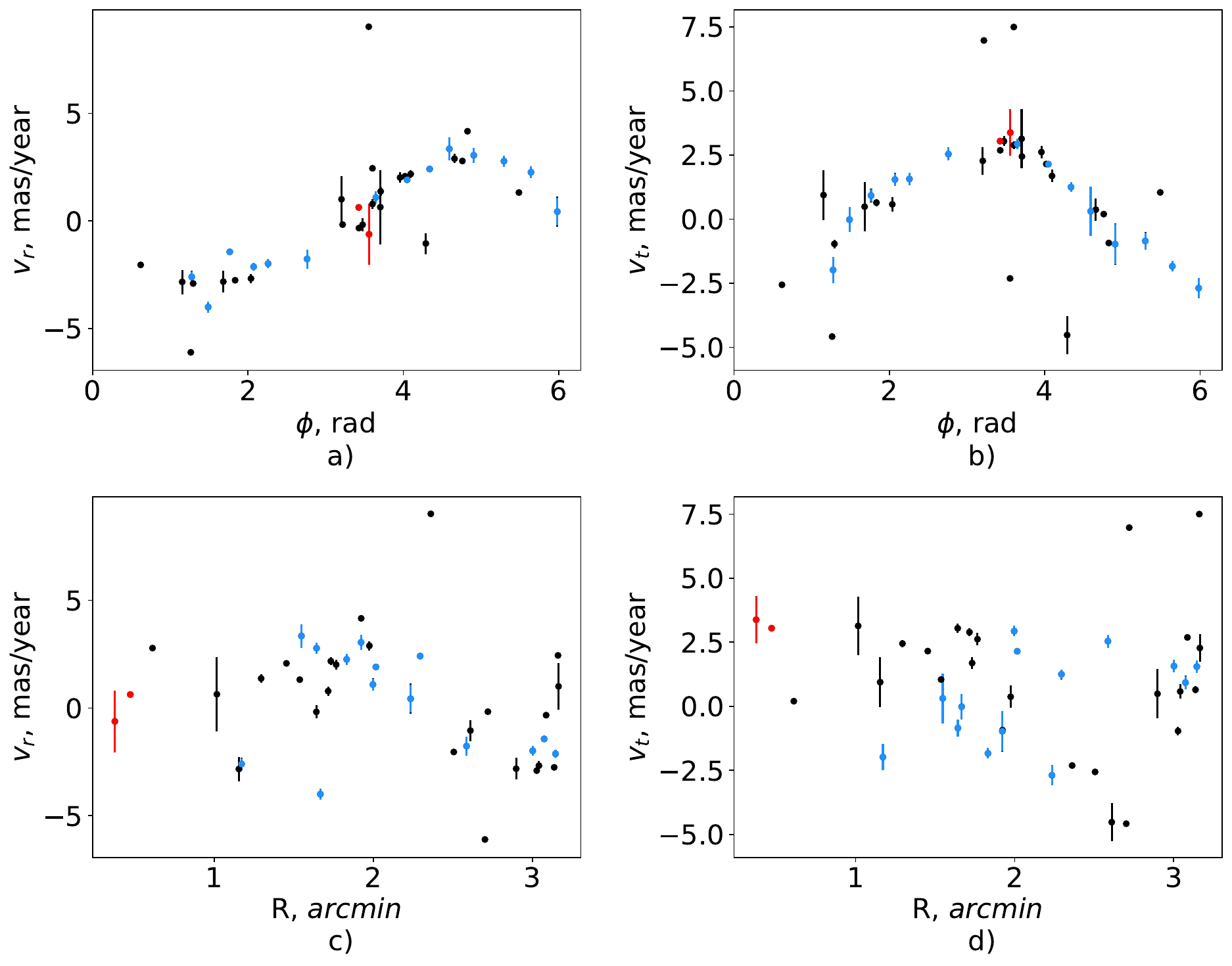}
      \caption{The same as Fig.\ref{FigVC10} for cluster 14 (S235~East1+East2).}
              \label{FigVC14}%
    \end{figure}

    Fig.\ref{FigVC10}, Fig.\ref{FigVC11}, and Fig.\ref{FigVC14} show the trend of `tangential component~-- positional angle', `radial component~-- positional angle', `tangential component~-- distance' and `radial component~-- distance' for clusters 10 (S235~North-West), 11 (S235~A-B-C), and 14 (S235~East1+East2), respectively.
    In the cluster 12 (S235~Central), the number of stars with known velocities is not enough to analyze the kinematics.

    From the inspection of these plots we can provide the following conclusions:
     \begin{enumerate}
     \item[$\bullet$] The trend of the velocity components versus positional angle for clusters 10 (S235~North-West) and 11 (S235~A-B-C) show a clear sinusoidal sequence with the same maximum and minimum amplitudes.\\
    \item[$\bullet$] The intersection of these sequences with the abscissa axis at angles 0 and $\pi$ radians for the radial component and $\pi/2$ and $3\pi/2$ radians for the tangential component indicates that the clusters move predominantly from north to south.\\
    \item[$\bullet$] The trend  of the velocity components versus distance from the cluster center does not show any dependence.\\
    \item[$\bullet$] For cluster 14 (S235~East1+East2),  plots of the velocity components versus positional angle show a deviation from the exact sinusoidal form (the left slope is larger than the right slope).
    This might be due to the small number of stars with known velocities in some parts of the cluster.
    Another possibility is the presence of the motion irregularities inside the cluster.
    We consider the close position of the apex of a majority of stars to the cluster center as the most probable reason of the deviation of the plot (in Fig. \ref{FigVC14}) from an exact sinusoidal for cluster 14 (S235~East1+East2).
    The trend of the velocity components versus distance for cluster 14 (S235~East1+East2) does not show the radial dependence similarly to the clusters 10 (S235~North-West) and 11 \textbf{(S235~A-B-C)}.\\
    \item[$\bullet$] We also note that the distribution of YSOs in clusters 10 (S235~North-West), 11 (S235~A-B-C), and 14 (S235~East1+East2) on the plots at Fig. \ref{FigVC10}, Fig. \ref{FigVC11}, and Fig. \ref{FigVC14} does not deviate from the distribution of other stars.
    \item[$\bullet$] There is also no dependence of the distribution of velocity components in clusters on the mass of stars (for more details on determining the mass of stars, see Section \ref{7.1}).
   \end{enumerate}

\section{Clusters' mass determination}\label{7}

    Since we know the probabilities of membership of stars in clusters 10 (S235~North-West), 11 (S235~A-B-C), 12 (S235~Central), and 14 (S235~East1+East2), we can attempt to estimate their masses.
    In order to achieve this, we use two methods: the first is the comparison with the evolutionary tracks of stars, and the second is a dynamic mass estimate.

\subsection{Comparison with evolutionary tracks}\label{7.1}

    We use evolutionary tracks of the Padova suite of models \citep {Bressan12} to determine the mass of clusters from the color-magnitude diagrams.
    We use evolutionary tracks with masses from 0.1 to 8 $M_{\odot}$ with  step of 0.1 $M_{\odot}$ for masses less than 1.0 $M_{\odot}$ and  step of 0.5 for masses greater than 1.0 $M_{\odot}$.
    We count the number of stars between two adjacent tracks and multiply this number by the average mass of the stars for which these tracks are built.
    For stars located to the right of the track for the lowest-mass star, we assume the masses of all stars equal to 0.1 $M_{\odot}$.
    Calculations are carried out for tracks built in the `H-K versus K' color magnitude diagram.

    The mass estimate obtained in this way may be underestimated.
    This can happen for several reasons:
    \begin{enumerate}
    \item[$\bullet$] Due to the high absorption in the region and the limitation of the catalog depth, most low-mass stars may not be visible.
    \item[$\bullet$] Some of the stars included in the cluster might not have a magnitude in all pass-bands and therefore they remain unaccounted for during selection by the UPMASK program.
    \item[$\bullet$] For some stars, the extinction has not been determined, since their parameter Q goes beyond the boundaries in which the non-reddened sequence is determined.
    \item[$\bullet$] We may erroneously consider a massive star to be low-mass one due to an underestimation of the absorption value.
    \end{enumerate}
    
    The outcome of the computation of the photometric mass $M_{ph}$ is summarized in Table \ref{TabClustMass}.

\begin{figure}
      \centering
      \includegraphics[width=15cm]{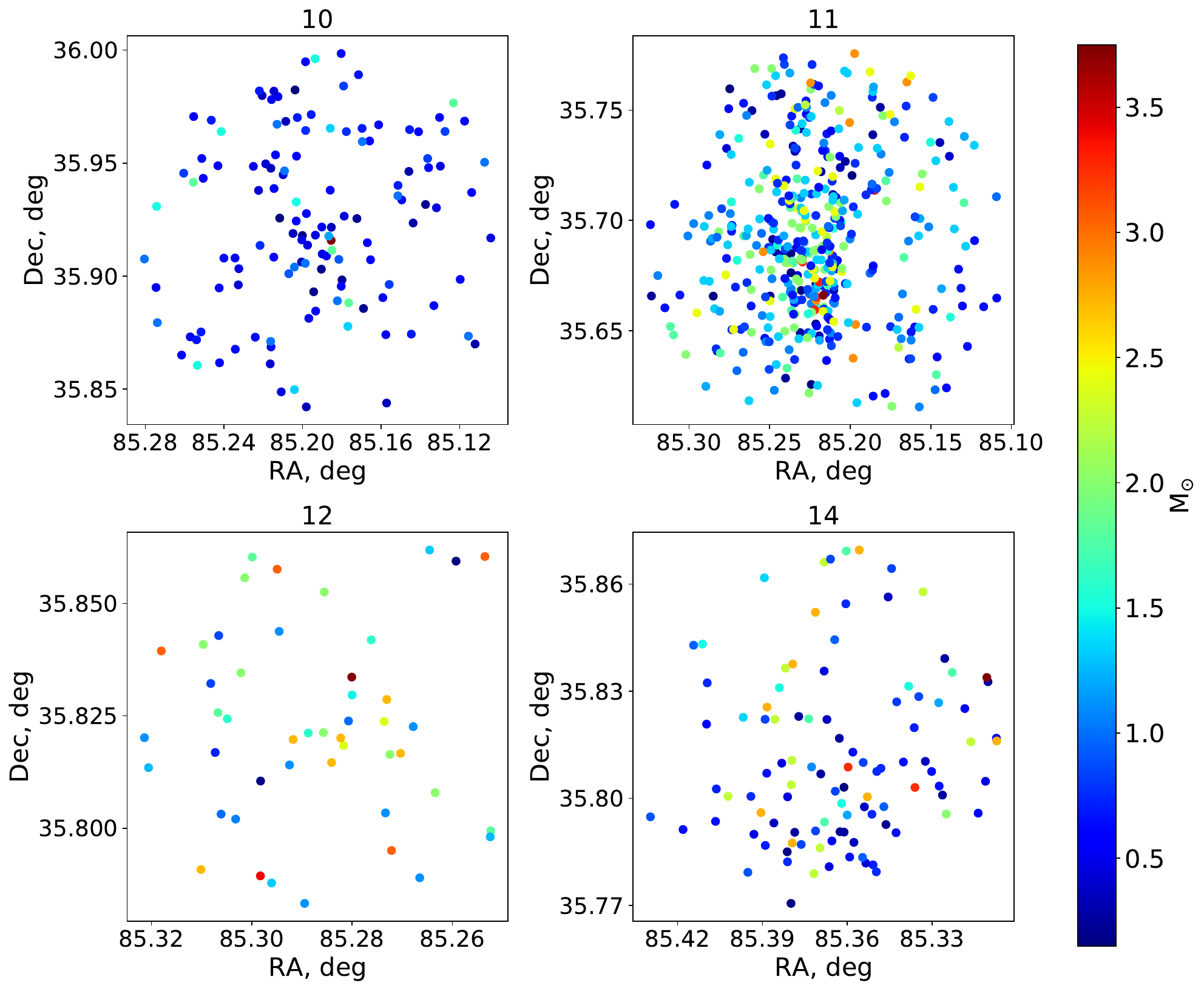}
      \caption{Map of the positions of stars in clusters 10 (S235~North-West), 11 (S235~A-B-C), 12 (S235~Central) and 14 (S235~East1+East2). 
      The color indicates the mass obtained by comparing the positions of stars on CMD with evolutionary tracks.}
              \label{FigMassMap}%
    \end{figure}

    Since this method allows us to determine in what mass range an individual star is located, we can study the distribution of stars of different masses within the cluster.
    Fig. \ref{FigMassMap} shows that the massive stars of cluster 11 (S235~A-B-C) are located mainly in the central part of the cluster.
    Clusters 12 (S235~Central) and 14 (S235~East1+East2) have a small number of massive stars, and they seem to be uniformly distributed.
    On the other hand, in cluster 10 (S235~North-West) massive stars are located mainly at the periphery.

\begin{figure}
      \centering
      \includegraphics[width=15cm]{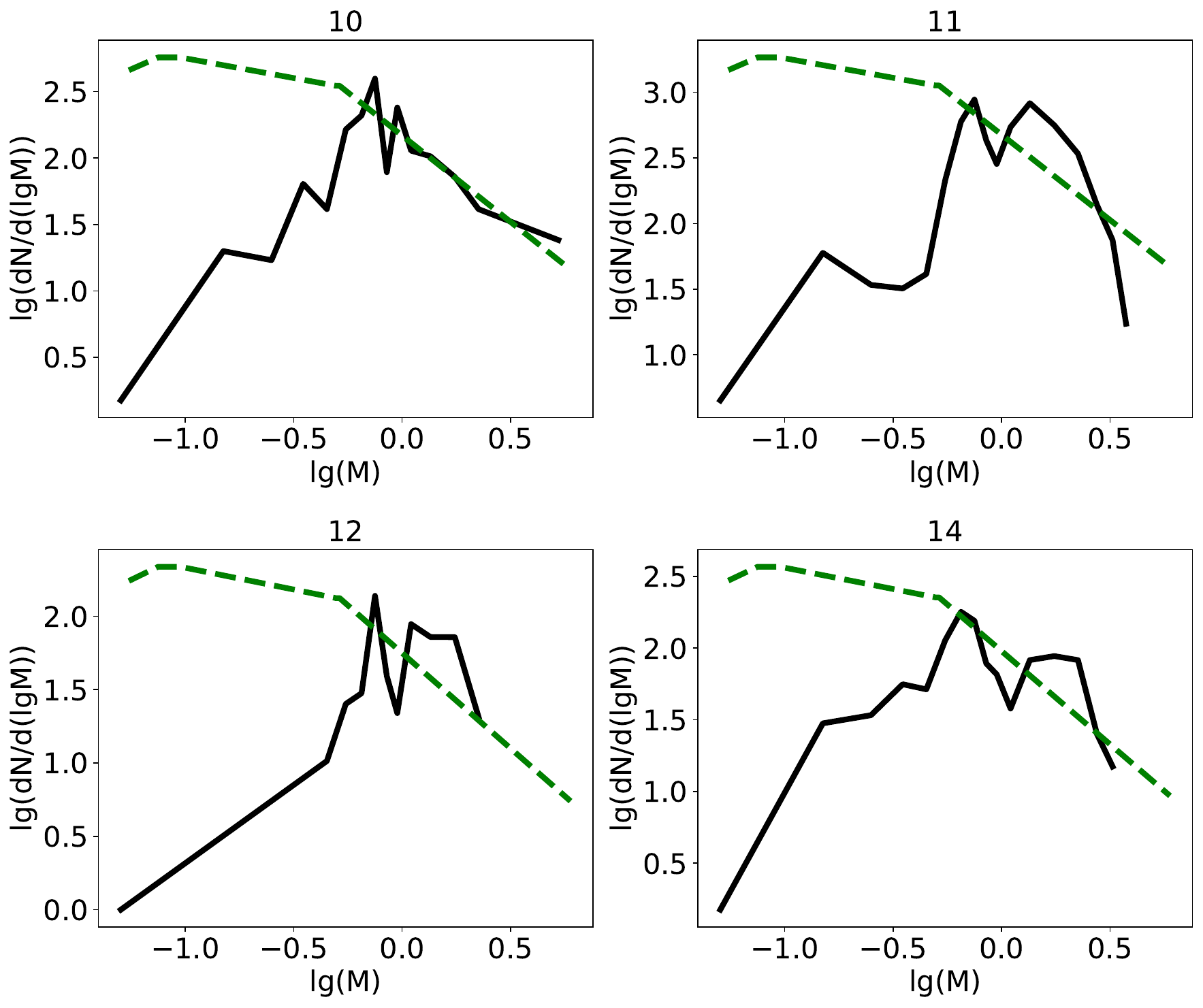}
      \caption{Mass functions of clusters 10 (S235~North-West), 11 (S235~A-B-C), 12  (S235~Central) and 14 (S235~East1+East2) (in black).
      The Kroupa initial  mass function is shown in green.}
              \label{FigMassFunc}%
    \end{figure}

    It is also possible to plot a cluster mass function (Fig. \ref{FigMassFunc}).
    To do this, we count cluster stars in mass intervals of width $0.1 \, \mathrm M_\odot$ (for $M<=1.0 \, \mathrm M_\odot$) or $0.5 \, \mathrm M_\odot$ (for $M>1.0 \, \mathrm M_\odot$) and plot $lg(M)$~versus $lg(dN/d(lg(M)))$.
    Here $lg(M)$ is the decimal logarithm of the mass of the center of the interval, $dN$ is the number of stars in the interval, $d(lg(M))$ is the width of the interval in which the stars were counted.
    Fig. \ref{FigMassFunc} shows that in the region of low masses (to the left of $lgM=-0.2$) there is a strong decrease in the number of stars.
    This reflects incompleteness in the catalog.
    The slope of the mass function to the right of $lgM=-0.2$ for clusters 10 (S235~North-West), 11 (S235~A-B-C) and 14 (S235~East1+East2) is close to the Kroupa parametric mass function \citep{Kroupa01}.
    In cluster 12 (S235~Central), due to the small number of stars, the mass function is characterized by a large change in the number of stars in neighbouring mass intervals.

    To estimate the total cluster mass, we use the assumption that the masses of stars in clusters are distributed according to the IMF.
    We also assume that we see all bright stars with masses greater than 0.6 $M_{\odot}$.
    Then, using the IMF, we can calculate the number of low-mass stars that we cannot see.
    Next we simply add the mass of these stars to the mass obtained from massive stars.
    We obtain the IMF from the power-law mass spectrum proposed by Kroupa \citep{Kroupa01}, which reads:

    \begin{equation}\label{Eq7.1.1}
    \xi(m)\propto m^{-\alpha},
    \end{equation}
    
    \noindent 
    where the exponent varies depending on the mass of the star:

    \begin{center}
    $\alpha=0.3,\enskip0.01\leqslant m <0.08\enskip M_{\odot},$
    \begin{equation}\label{Eq7.1.2}
    \alpha=1.3,\enskip0.08\leqslant m <0.50\enskip M_{\odot},
    \end{equation}
    $\alpha=2.3,\enskip0.5\leqslant m\enskip M_{\odot}.$
    \end{center}
    
    \noindent 
    IMF is also given by Eq. \ref{Eq7.1.1}, with exponent $\alpha-1$.
    To count stars, it is necessary to find a normalization coefficient, individual for each cluster.
    To determine the coefficient, we take stars with masses greater than 0.6 $M_{\odot}$.
    Using the curve\_fit method of the $\texttt{SciPy}$ library of $\texttt{Python}$, we combine a segment of the Kroupa mass function with an exponent $\alpha=2.3$ with the selected segment of the cluster mass function.
    Next, using the normalized function, we determine the number of stars with masses less than 0.6 $M_{\odot}$.
    We convert the $lg(dN/d(lg(M)))$ values of the Kroupa mass function at points with masses 0.15, 0.25, 0.35, 0.45 and 0.55 $M_{\odot}$ into the number of stars in 0.1 $M_{\odot}$ intervals.
    To get the total number of stars in a cluster, we add this number of stars to the number of stars with masses greater than 0.6 $M_{\odot}$.
    We calculate the total mass of stars with masses less than 0.6 $M_{\odot}$ as the sum of the products of the number of stars in the mass interval and the average mass of the interval.
    To estimate the total mass of the cluster, we add the mass found from visible stars with $M\geq0.6 M_{\odot}$ with the mass determined from the Kroupa function for stars with $M<0.6 M_{\odot}$.
    The resulting cluster masses ($M_{Kr}$) and number of stars ($N_{Kr}$) are given in Table \ref{TabClustMass}. 
    The masses and number of stars calculated from the Kroupa IMF are almost 2 and 3 times, respectively, larger than the values obtained from visible stars.

    \begin{table}[h]
    \caption{Mass of the clusters in $M_\odot$ and the velocity dispersion of cluster stars.}
    \label{TabClustMass}      
    \centering                          
    \begin{tabular}{cccccc}       
    \hline                 
    № & $M_{ph}$ & $M_{Kr}$ & $N_{Kr}$ & $M_d$ & $\sigma^2$, (km/s)${}^2$ \\ 
    \hline                        
       10 & $113\pm17$ & $209\pm36$ & $485\pm84$ & $186000\pm10000$ & $213.3\pm1.2$ \\  
       11 & $530\pm79$ & $873\pm99$ & $1664\pm157$ & $172000\pm5000$ & $154.9\pm0.4$ \\
       12 & $56\pm10$ & $91\pm15$ & $189\pm44$ & $59000\pm5000$ & $97.3\pm0.6$ \\
       14 & $95\pm15$ & $156\pm30$ & $320\pm64$ & $31000\pm2000$ & $53.3\pm0.3$ \\

    \hline                                   
    \end{tabular}
    \end{table}

    Previously,  masses of these clusters were calculated by \citet{Camargo11} and \citet{Chen24}.
    The estimates obtained by \citet{Chen24} are compatible with masses of clusters 12 (S235~Central) and 14 (S235~East1+East2), that we obtained from visible stars, while they are approximately two and ten times lower for clusters 11 (S235~A-B-C) and 10 (S235~North-West), respectively.
    \citet{Camargo11} determined the masses only for two clusters considered in our work.
    The mass estimate for cluster 12 (S235~Central) is in agreement with ours, whereas for cluster 14 (S235~East1+East2) it is approximately half as much.
    The discrepancies are most likely due to differences in methods for determining whether stars belong to clusters, the size of the regions of these clusters, and the use of different catalogs (UKIDSS gets fainter than 2MASS).

\subsection{Dynamic mass of clusters}\label{7.2}

    A dynamic estimate of the cluster mass can be obtained based on the assumption of virial equilibrium and cluster homogeneity using the following equation \citep {Seleznev17}:

    \begin{equation}\label{Eq7.2.1}
    M=\frac {5R\sigma^2}{3G}.
    \end{equation}

    Here M is the cluster mass, $\sigma^2$ is the cluster velocity dispersion, R is the cluster radius, G is the gravitational constant.
    By velocity dispersion $\sigma^2$ we mean the mean square of the star's velocity relative to the centroid of the velocities of the cluster stars \citep {Seleznev17}.
    
    The velocity dispersion can be found from the dispersion of proper motions of the stars ($\sigma_{pmRA}$ and $\sigma_{pmDec}$), assuming that the distribution of radial velocities is the same as the distribution of velocities in one tangential direction:

    \begin{equation}\label{Eq7.2.2}
    \sigma^2=1.5(\sigma^{2}_{pmRA}+\sigma^{2}_{pmDec}).
    \end{equation}
    
    However, the estimate of dispersion obtained by Eq. \ref{Eq7.2.2} is an over-estimate, since it contains errors in measured proper motions.
    To determine the true velocity dispersion from the observed one, we use the method proposed by \citet{Kulesh24}.
    
    The method stands on the assumption that the observed distribution of a random variable ($\sigma_{obs}$) is a convolution of the true distribution ($\sigma_{tr}$) and the distribution of the error of this variable ($\sigma_{err}$).
    If we assume that all distributions are Gaussian and described by the function:

    \begin{equation}\label{Eq7.2.3}
    f(x)=\frac{1}{\sigma_0\sqrt{2\pi}}e^{-\frac{(x-m)^2}{2\sigma_0^2}}
    \end{equation}
    
    \noindent (where $\sigma_0$ is the standard deviation, $m$ is the mathematical expectation), then

    \begin{equation}\label{Eq7.2.4}
    \sigma^{2}_{obs}=\sigma^{2}_{tr}+\sigma^{2}_{err}+\sigma^{2}_{KDE}.
    \end{equation}
    
    \noindent Here $\sigma^{2}_{KDE}$ takes into account the contribution of the distribution broadening due to the use of the KDE method when constructing this distribution.
    Thus, the true dispersion of the proper motion in declination or right ascension is calculated using the equation:
    
    \begin{equation}\label{Eq7.2.5}
    \sigma^{2}_{tr}=\sigma^{2}_{obs}-\sigma^{2}_{err}-\sigma^{2}_{KDE}.
    \end{equation}
    
    Following \citet {Kulesh24} we assume a normal distribution of the velocity errors, and, to find the dispersion of their distribution, we use the distribution of their squares: 

    \begin{equation}\label{Eq7.2.6}
    g(x)=\frac{1}{2\sigma_0\sqrt{2\pi x}}e^{-\frac{(\sqrt{x}-m)^2}{2\sigma_0^2}}.
    \end{equation}

    \noindent Here the notation is similar to that in Eq. \ref{Eq7.2.3}.
    We cannot use a Gaussian to find the dispersion of velocity errors since we only know their absolute values.

\begin{figure}
      \centering
      \includegraphics[width=15cm]{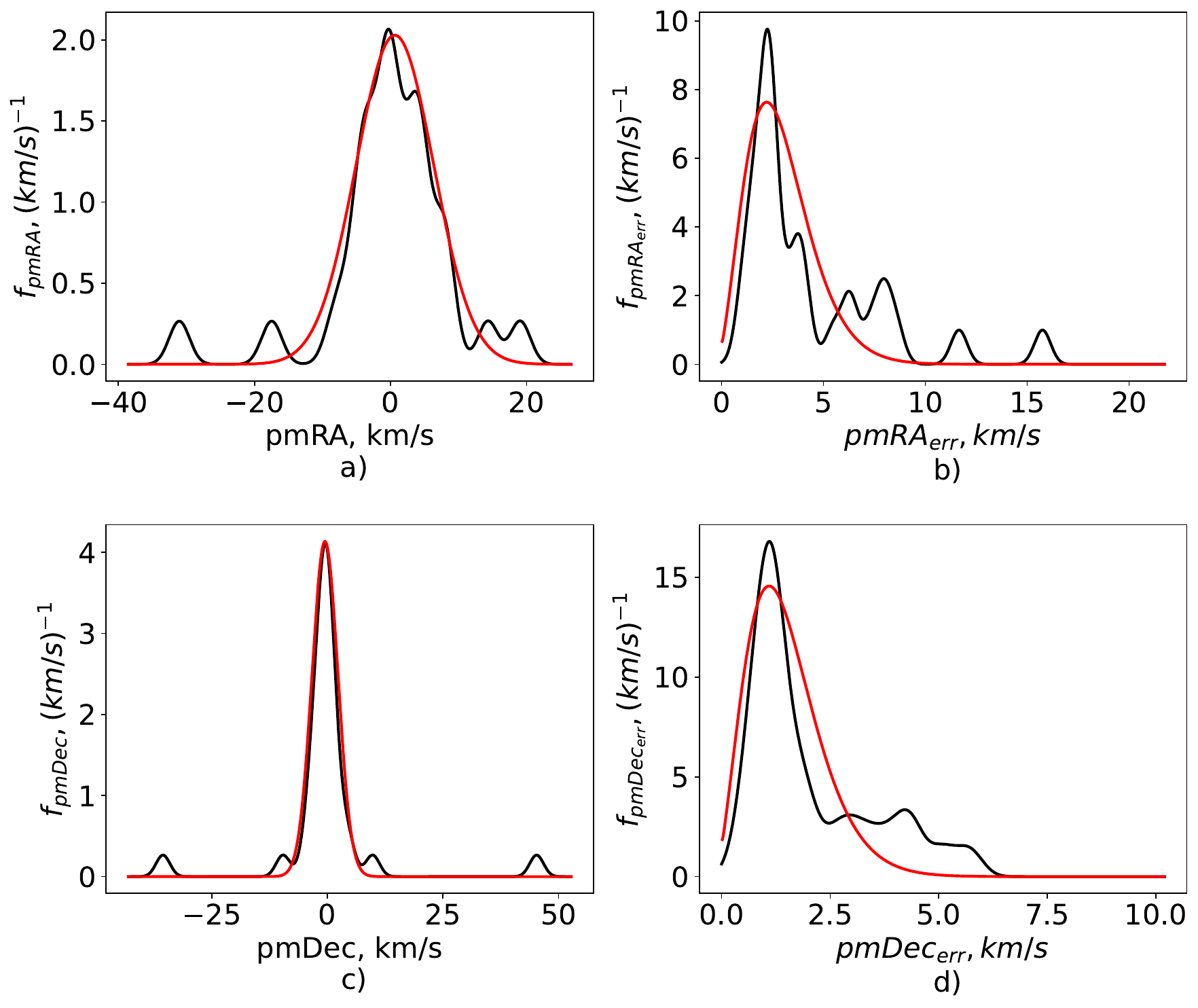}
      \caption{Distribution of proper motions of stars and their errors in cluster 14 (S235~East1+East2) (black).
      Functions that approximate distributions are shown in red.}
              \label{FigDispPM}%
    \end{figure}

    To obtain the distribution of velocities and their errors, we use the KDE method with the kernel given by Eq. \ref{Eq7.2.3}.
    The values of the kernel half-widths ($\sigma_{KDE}$) for constructing the distribution of velocities and their errors were selected manually, so that the peak of the distribution was clearly distinguished.
    To construct the distribution of errors in proper motions in declination, smaller values were used than for errors in right ascension.
    It is worth remembering that the distribution of squared errors is defined in the interval $(0,\infty)$.
    We approximate the resulting distributions with functions given by the Eq. \ref{Eq7.2.3} and \ref{Eq7.2.6} multiplied by the number of stars used to construct the distribution.
    To find the parameters of the approximating functions, we use the curve\_fit() function of the \texttt{Python SciPy} library.
    The standard deviations obtained by the program are the observed dispersions of the quantities under consideration.
    An example of the distribution of proper motions by declination and right ascension, errors of these values and functions approximating them is shown in Fig. \ref{FigDispPM} for cluster 14 (S235~East1+East2).
    Next, we calculate the dispersion using Eq. \ref{Eq7.2.5} and \ref{Eq7.2.2}.
    To find the dynamic masses of clusters, we substitute the resulting dispersions and radii of the clusters in which the star memberships are determined into the equation \ref{Eq7.2.1}.
    Estimates of the three-dimensional dispersion ($\sigma^2$) obtained from proper motions using the formula \ref{Eq7.2.2} and cluster dynamical masses ($M_d$) are given in the Table \ref{TabClustMass}.

    The mass estimates obtained using the Eq. \ref{Eq7.2.1} are several orders of magnitude larger than the one determined from evolutionary tracks.
    This may be due to the unaccounted mutual interaction between stars and gas, and the fact that clusters are most probably be non-stationary (out of virial equilibrium).
    As shown in \citet{Danilov24}, the use of the virial theorem for an isolated system to determine the mass of a real cluster can lead to an overestimation of the cluster mass by more than 2 times.
    It is also worth underlying that the stars in the region under consideration have high extinction and large magnitudes in the optical band, which complicates the determination of their proper motions.
    For this reason, estimates of the velocity dispersion of cluster stars obtained from the proper motions of the Gaia catalog may be overestimated.
    The errors of the proper motions given in the catalog may also be underestimated. In turn, one can expect a bias towards larger dynamical masses.

\section{Mass of gas}\label{8}

    To estimate the mass of molecular gas in the region, we use observational data on emissions in the ${}^{12}$CO(1-0) and ${}^{13}$CO(1-0) lines.
    They were extracted from the Extended Outer Galaxy Survey (E-OGS, \citet{Brunt04}), which was carried out on the 13.7-m telescope of the Five College Radio Astronomy Observatory (FCRAO) with the Second Quabbin Observatory Imaging Array (SEQUOIA), a 32-pixel focal receiver.
    The telescope beam size is 45" for ${}^{12}$CO(1–0) and 46" for ${}^{13}$CO(1–0).
    The observed frequency was set to 115.27120 GHz for the ${}^{12}$CO(1–0) line and 110.20135 GHz for the ${}^{13}$CO(1–0) line.
    The spatial step was $22.5"$, which is two times smaller than the half-power beam-width (HPBW).
    Spectral resolution was 0.127 km/s for  the ${}^{12}$CO(1–-0) line and 0.133 km/s for the ${}^{13}$CO(1--0) line.
    The noise level $\sigma_{T_{mb}}$ for ${}^{12}$CO(1--0) was 1.1 K and for ${}^{13}$CO(1--0) was 0.63 K on the main-beam temperature scale.
    Observations of the G174+2.5 region were carried out in January 2000.
    The line emission map of both CO isotopes covers an area of size $150'\times150'$ with the center $RA=85.0^{\circ}$ and $Dec=36.12^{\circ}$.
    To process data in CO lines and calculate physical parameters, we used the MIRIAD package \citep{Sault95}.

    Emission maps ${}^{12}$CO(1--0) and ${}^{13}$CO(1--0) were converted into column density maps similar to the method described in \citet{Ladeyschikov21}.
    We used the gas column density equation using the LTE approach described by \citet{Mangum15}:
    
    \begin{equation}\label{Eq8.1}
    N=\frac{\int\tau}{1-\mathrm{exp}(\int\tau)}\frac{3\mathrm h}{8\pi^3\mathrm S\mu^2}\frac{\mathrm{Q_{rot}}}{\mathrm {g_J}}\frac{\mathrm{exp}(\frac{\mathrm{E_u}}{k\mathrm{T_{ex}}})}{\mathrm{exp}(\frac{\mathrm h\nu}{k/\mathrm{T_{ex}}})-1}\frac{1}{f}\int\tau \mathrm d\nu ,
    \end{equation}

    \noindent where $\mathrm{S = J^2/([2J + 1])}$, $\mathrm{g_J=2J+1}$ is the statistical weight, and $\mathrm{Q_{rot}}\simeq (k\mathrm{T_{ex}/h/B_{rot}})+1/3$ is the partition function for linear molecules, $f = 1$ filling factor we adopted, with $T_{ex}$ being the excitation temperature of the rotational line.
    The constants $\mathrm{B_{rot}}$, $\mathrm{E_u}$ and $\mu$ were taken from the Cologne Database for Molecular Spectroscopy \citep{Muller01}.
    For ${}^{12}$CO(1-0) they are $\mathrm{B_{rot}}=57.6$ GHz, $\mathrm{E_u}=5.53$ K and $\mu=0.11\cdot10^{-18}$ esu, for ${}^{13}$CO(1-0) -- $\mathrm{B_{rot}}=55.1$ GHz, $\mathrm{E_u}=5.29$ K, and $\mu=0.11\cdot10^{-18}$ esu.

    The line-integrated optical depths of the CO lines were estimated from the ratio of the integrated over the line profiles main-beam brightness temperatures of the optically thick ${}^{12}$CO(1--0) and optically thin ${}^{13}$CO(1--0) lines using the equation:

    \begin{equation}\label{Eq8.2}
    \frac{\int\mathrm{T_{mb {}^{13}CO}}}{\int\mathrm{T_{mb {}^{12}CO}}}=\frac{1-e^{-\int\tau/R}}{1-e^{-\int\tau}},
    \end{equation}

   \noindent where the isotopic ratio $R=80$.

    $T_{\rm ex}$ was determined using the optically thick $^{12}$CO(1--0) line according to the equation:

    \begin{equation}\label{Eq8.3}
    \mathrm{T_{ex}}=\mathrm{T_0}/\mathrm{ln}(1+\frac{\mathrm{T_0}}{\mathrm{T_B^{12}+T_0/(e^{T_0/T_{bg}}-1)}}),
    \end{equation}

    \noindent where $\mathrm{T_B^{12}}$ is the peak brightness temperature of ${}^{12}$CO, $\mathrm{T_{bg}=2.7}$ K and $\mathrm{T_0= h}\nu/k$.

    To convert $N$(CO) to $N$(H$_2$), we use the abundance ratio $\mathrm{[CO]/[H_2]}=8\cdot10^{-5}$ \citep{Simon01}.
    This value is supported by abundance of atomic carbon $[C]/[H]$ \citep{Cardelli+1993,Savage&Sembach1996,Sofia+2004} by observations of interstellar absorption lines.
      
    Finally, the mass of molecular gas in every pixel of the map is calculated from the column density of molecular hydrogen using the equation 
    
    \begin{equation}\label{Eq8.4}
    M=N({\rm H_2})\mu_{H_2} m_{\rm H} A,
    \end{equation}
    
    \noindent where $\mu_{H_2}= 2.8$ is the  mean weight of molecular interstellar gas 
    , $m_{\rm H}$ is the mass of an hydrogen atom, and A is the area of the pixel at the plane of the sky at a distance equal to the photometric distance to the cluster (Table \ref{TabClustPar}) in $\mathrm{cm}^2$. 
    
    We estimate the gas mass in each cluster region using the integration of the cluster area in the mass-in-pixel maps.
    Errors of masses, column densities, and temperatures were estimated using the \texttt{UNCERTAINTIES} package of \texttt{python} and the root mean square (rms) maps of ${}^{12}$CO(1--0) and ${}^{13}$CO(1--0).
    The resulting gas masses ($M_g$) are given in the table \ref{TabClustEn}.

   The above method allows us to estimate the mass of gas located along the line of sight in the region under consideration.
   This means that we account for the mass of gas not only inside, but also behind and in front of the cluster.
   Assuming that this gas was swept out of the cluster during evolution, but was in the region of the cluster at the time of its formation, we estimate the efficiency of star formation using the following equation:

    \begin{equation}\label{Eq8.5}
    SFE=\frac{M_{cl}}{M_g+M_{cl}},
    \end{equation}

    \noindent where $M_g$ is the mass of gas in the region with the cluster, $M_{cl}$ is the mass of clusters, which we take as the estimate calculated using the Kroupa IMF ($M_{Kr}$ in Table \ref{TabClustMass}).
    See the Table \ref{TabClustEn} for the obtained estimates.

    Typically, estimates of star formation efficiency are in the range 0.03 - 0.42 \citep[see, for example,][]{Lada03,Megeath16,Lada92}.
    Our values range from 0.04 to 0.17, which are quite close to the observed averages.

    Differences in estimated star formation efficiency between regions may be due to both variations in the mode of star formation and differences in age.
    The older the cluster, the more parent gas has managed to leave it.
    Accordingly, in more evolved clusters, we expect to obtain a greater estimate of the efficiency of star formation due to the smaller amount of remaining gas than in younger clusters.
    During induced star formation, it is expected that gas from the cluster region will be removed not only by the radiation of the stars, but also by the pressure of the expanding HII shell.
    Thus, the estimate of the efficiency of star formation, due to the smaller amount of gas remaining in the cluster, may be higher than during spontaneous star formation.
    However, it is worth remembering that due to possible features of the history of the evolution of the gas component of a particular region that are unknown to us, the assumptions voiced above may not be fulfilled.

    The highest SFE values are associated with clusters 10 (S235~North-West) and 11 (S235~A-B-C).
    Cluster 10 (S235~North-West) is located near the boundary of the HII region, so we can assume that star formation in it is most likely induced.
    Cluster 11 (S235~A-B-C) is located at a considerable distance from the expanding envelope and therefore its formation is probably unrelated to the expansion of S235.
    At the same time, in cluster 10 (S235~North-West) a lower absorption value is observed compared to 11 (S235~A-B-C).
    This may indicate that some gas has been removed from region 10 (S235~North-West) and, as a result, leads to an overestimation of the SFE in the cluster.
    The lowest efficiencies are observed in clusters 12 (S235~Central) and 14 (S235~East1+East2) with induced star formation, which was triggered by the expansion of the HII region S235 \citep{Kirsanova08, Kirsanova14}.
    As can be seen, the smallest SFE in the region under consideration is most likely associated with regions of induced star formation.
    However, the differences in SFE estimates between clusters are not so significant as to indicate in this case that there is a strict relationship between SFE and the process of the cluster formation.
    Moreover, in the clusters under consideration, the efficiency does not appear to depend on their age.
    In the studied clusters, the ages are estimated to be 3 (10 (S235~North-West),11 (S235~A-B-C),14 (S235~East1+East2)) and 5 (12 (S235~Central) million years \citep{Camargo11}.
    In younger clusters, there is a scattering of SFE estimates, and in more evolved clusters, contrary to expectations, SFE is lower.

    \textbf{The obtained SFE estimates are approximate, since this value is influenced by many factors.
    So we can only see the gas that is in the cluster now.
    However, gas is swept out of the cluster during the process of evolution.
    In addition, a significant portion of the gas is ionized and not visible in molecular lines.
    All this leads to an underestimation of the gas mass and an increase of SFE.
    Estimates of the cluster's mass may also contain inaccuracies.
    Some of the low-luminosity stars are not visible due to photometric limitations and significant absorption in the region.
    Although we try to take into account the mass and number of low-mass stars using the Kroupa mass function, these estimates may still not correspond to the actual mass of the cluster.
    }

    \begin{table}[h]
    \caption{Mass of the gas in $M_\odot$ and energy of clusters.}
    \label{TabClustEn}      
    \centering                          
    \begin{tabular}{ccccc}       
    \hline                 
    № & $M_g$ & SFE & E $M_{\odot}\cdot s^{-2}\cdot km^{2}$ & $E_{all cloud}$ $M_{\odot}\cdot s^{-2}\cdot km^{2}$\\ 
    \hline 
       10 & $1260\pm20$ & $0.14\pm0.02$ & $21100\pm3800$ & $-89400\pm6000$ \\  
       11 & $4360\pm30$ & $0.17\pm0.02$ & $47300\pm7800$ & $-54600\pm9100$ \\
       12 & $2180\pm20$ & $0.04\pm0.01$ & $-1900\pm1100$ & $-105600\pm4600$ \\
       14 & $2150\pm20$ & $0.07\pm0.01$ & $-3000\pm1200$ & $-106800\pm4600$ \\

    \hline                                   
    \end{tabular}
    \end{table}

\section{Dynamical state of the clusters}\label{9}

    Recently, \citet{Danilov24} considered the problem of the dynamics of an embedded cluster inside the parental gas-dust cloud within the framework of the gross dynamic approach.
    We use his results to evaluate the total energy of the clusters with the aim to estimate their possible future.
    If we determine the sign of the total energy, we can infer whether the system is gravitationally bound or not.

    The derivation of the total energy equations is based on the method of dynamically isolated k groups (clusters) of stars, developed in \citet{Danilov84}. 
    It allows us to describe the dynamics of spherically symmetric clusters of stars that are nonstationary in a regular field. 
    In such systems, changes in the total energies of groups of stars are caused by the interactions of these groups with each other and their interactions with the variable force field of the system. 
    This method is also applicable to describe cluster-gas systems.

    The total energy is calculated on the assumption that the system consists of two components~-- stars (1) and gas (2).
    It is assumed that the radius of the stellar component is less than or equal to the radius of the gas component. 
    In the method used, both subsystems are considered as homogeneous gravitating spheres with coinciding centers of mass.
    For each of these subsystems we need to know the mass, radius and velocity dispersion.
    To estimate the kinetic energy of the stellar and gaseous subsystems, we used the equation

    \begin{equation}\label{Eq9.1}
    T_i=\frac{M_i\sigma_i^2}{2}.
    \end{equation}

    \noindent Here $M_i$ is the mass of the cluster or gas, $\sigma_i^2$ is the velocity dispersion of the cluster stars or the dispersion of non-thermal gas motion.
    The potential energy of the system consists of the potential energy of each of the subsystems (first term in every equation) and the energy of their interaction (second term):

    \begin{equation}\label{Eq9.2}
    \Omega_1=-\frac{3GM_1^2}{5R_1}-\frac{3GM_1M_2}{2R_2}\left(1-\frac{R_1^2}{5R_2^2}\right),
    \end{equation}
    \begin{equation}\label{Eq9.3}
    \Omega_2=-\frac{3GM_2^2}{5R_2}-\frac{3GM_1M_2}{2R_2}\left(1-\frac{R_1^2}{5R_2^2}\right),
    \end{equation}

    \noindent where $M_i$ are masses, $R_i$ are radii of subsystems, $G$ is gravitational constant.
    Thus, the total energy of the system results:
    
    \begin{equation}\label{Eq9.4}
    E=E_1+E_2- \Omega_{12}=T_1+\Omega_1+T_2+\Omega_2 - \Omega_{12},
    \end{equation}

    \noindent where $\Omega_{12}$ is the potential energy of interaction of the stellar subsystem with the gas subsystem

     \begin{equation}\label{Eq9.5}
    \Omega_{12}=-\frac{3GM_1M_2}{2R_2}\left(1-\frac{R_1^2}{5R_2^2}\right).
    \end{equation}

    It is worth highlighting that the calculations in \citet{Danilov24} were carried out for the case when all subsystems are spherically symmetric and their centers coincide. 
    The clusters we are considering, however, have a more elongated shape or, like cluster 11 (S235~A-B-C), a hierarchical structure.
    Also, when estimating the total energy of the system taking into account the gas mass of the entire cloud, it is obvious that the geometric centers of the cloud and clusters will not coincide.
    For such complex shapes, energy calculations are not possible, so to obtain approximate energy estimates we use equations for the spherically symmetric case.
    
    To calculate the total energy we take the following values.
    The radius of both the stellar and gas components is taken as the radius in which the probabilities of membership of stars in clusters were determined (see Section \ref{4.2}).
    The velocity dispersion of the cluster stars is defined in Section \ref{7.2}.
    For the total mass of the cluster we use the mass obtained using the Kroupa's mass function (see Section \ref{7.1}), while
    the mass of the gas in each region is defined in Section \ref{8}.
    We considered the dispersion of gas velocities in the regions of clusters to be equal to the average dispersion of the non-thermal motion of the clumps, and the values are taken from \citet{Shimoikura18}.
    Note that by dispersion we mean the value of $\sigma^2$, and \citet{Shimoikura18} means the value of $\sigma$, so these values were squared.
    Plugging these quantities into Eq. \ref{Eq9.1}, \ref{Eq9.2} and \ref{Eq9.3}, we recover the potential and kinetic energies of the stellar and gaseous subsystems.
    Hence, substituting them into equation Eq. \ref{Eq9.4}, we obtain an estimate of the total energy of the systems.
    The calculated values E are shown in  Table \ref{TabClustEn}.

    The total energy in clusters 12 (S235~Central) and 14 (S235~East1+East2) is negative, which may indicate that the cloud-cluster system is gravitationally bound.
    However, the energies of star clusters (without the gas) in all cases are larger than zero.
    One can argue that they are gravitationally unbound and in the process of further evolution and removal of the gas from the region will eventually disperse or lose a significant number of stars.
    On the other hand, as discussed above (see Section \ref{7.2}), the velocity dispersion of cluster stars is most probably overestimated.
    Accordingly, the estimates of the kinetic energy of clusters and the total energies would also be overestimated.
    It is also worth noting that the velocity dispersions of cluster stars greatly exceed the velocity dispersions of gas.
    Such an excess is observed both for the gas dispersion values obtained by \citet{Shimoikura18} and for those estimated from the gas velocities obtained in the work of \citet{Kirsanova14}.

    Additionally, the energy of the systems was calculated under the assumption that the entire cloud exerts a gravitational influence on the cluster.
    The mass of the cloud in this case was determined as in Section \ref{8} in a region with center $RA=85.255^{\circ}$ and $Dec=35.76^{\circ}$, radius 18’ and at an average distance of 1.86~kpc.
    The resulting mass of the gas: $M_g=21560\pm70$ $M_{\odot}$.
    The calculated total energy values for this case ($E_{all cloud}$) also are given in Table \ref{TabClustEn}.
    The total energy of cluster-gas systems is negative in all cases - most likely the systems are gravitationally bound.


\section{Conclusions and Discussion}\label{6}

    In this work, we have studied the wide vicinity of the star formation region located in GMC G174+2.5 using the UKIDSS catalog to find previously unknown clusters and determine the parameters of all clusters.
    
   \begin{enumerate}
      \item  We found 14 clusters on the surface density map, 6 of which are previously unknown embedded clusters.
      All detected clusters are located in the molecular gas filaments.
      In the surface density map of the S235~A-B-C cluster region, we also found a number of local density maxima.
      We consider these density maxima to be sub-clusters of S235~A-B-C rather than separate clusters.
      \item For all clusters, we estimated the radii and field density levels in the cluster areas using the KDE method.
      Also, we have counted the cluster star numbers and 
      for 4 clusters estimated the membership probabilities of stars using the UPMASK package.
      \item The individual reddening of stars in the region was determined using the NICEST method and Q-method at the base of the UKIDSS photometry.
      Using NICEST data, we built the map of the absorption distribution in the region.
      The areas of increased absorption coincide with the filaments of molecular gas identified by the GetFilaments algorithm.
      \item The estimates of distance to clusters 10 (S235~North-West), 11 (S235~A-B-C), 12 (S235~Central) and 14 (S235~East1+East2) are in the range from 1.7 to 2.2 kpc.
      It indicates that these clusters belong to the studied star formation region.
      \item We use the proper motion data from Gaia DR3 to estimate the average velocities of stars in clusters.
      Also we study the internal kinematics of stars in some clusters: clusters 10 (S235~North-West) and 11 (S235~A-B-C appear to move as a single units (with respect to their centers) without noticeable internal motions of stars, while cluster 14 (S235~East1+East2) may have irregularities of stellar movement.
      \item The masses of four clusters were determined in two ways.
      The dynamic method gives greatly overestimated values, which is due to the lack of consideration of the gas component and large errors in determining proper motions.
      On the contrary, masses determined photometrically may be underestimated. 
      This is possible because stars with large magnitudes are not visible due to limitations in the completeness of the catalog and high extinction in the region.
      The total mass of the clusters was obtained using Kroupa's initial mass function.
      Additionally, based on the emission data of the lines of the ${}^{12}$CO(1-0) and ${}^{13}$CO(1-0), the gas masses in the areas of clusters were determined.
      \item We estimated star formation efficiency at the base of the obtained masses of gas and cluster stars.
      SFE ranges from 0.04 to 0.17, that corresponds to the values of star formation efficiency in the Milky Way known from the literature.
      The lowest estimates are observed in regions 12 (S235~Central) and 14 (S235~East1+East2) with induced star formation.
      The SFE estimate in the 10 (S235~North-West) cluster (also with induced star formation) may be exaggerated due to an underrating of the gas mass.
      Thus, in the region under consideration, induced star formation is characterized by lower SFE values, while spontaneous star formation (cluster 11 (S235~A-B-C)) is characterized by slightly higher values.
      However, the differences in the SFE estimates between clusters are not significant enough to confidently assert the connection between SFE and the cluster formation process.
      \item The total energies of four clusters were estimated considering both stellar and gas components.
      The gravitational boundness of the systems under consideration depends on the region for which the gas mass is estimated.
      Thus, when we take into account the mass of the entire cloud, all four cluster-gas systems appear to be gravitationally bound.
      When we take into account the gas only in the cluster regions, only clusters 12 \textbf{(S235~Central)} and 14 \textbf{(S235~East1+East2)} are probably gravitationally bound.
      
   \end{enumerate}

   It should be noted that both star counts and the analysis of the dynamical state of the clusters in our paper are based on an assumption of a spherical symmetry.
   However, the real structure and formation path are expected to differ from one cluster to another.
   Star formation in the clusters S235~North-West (No. 10 in Table 1), S235~Central (No. 12 in Table 1), S235~East1,East2 (No. 14 in Table 1) was most probably induced by an expanding HII region Sh2-235 \citep{Kirsanova08,Kirsanova14}.
   Consequently, these clusters have a flattened form.
   Star formation in the cluster S235~A-B-C (No. 11 in Table 1) was on the other hand spontaneous.
   Nevertheless, this cluster has a non-uniform structure and consists of some sub-clusters (Fig. 4d).
   
   Besides, the formation of some clusters in the envelope complicates the determination of the gas mass.
   A remarkable part of the gas there is ionized and is not visible in the molecular lines.
   Thus, our investigation can be considered as an attempt to use simple models for a complex star formation region.

   It is interesting that star formation region G174+2.5 seems to be a part of a larger structure containing HII regions, Supernova remnants, and O-B stars.
   This complex has a dumbbell structure with the approximate center at (173,0) and the larger axis perpendicular to the Galactic plane (see Fig.5 in \citet{Kang+2012}).
   This possibly indicates an even more tangled star formation history.
   However, to get a more detailed picture of the cluster kinematics and dynamics in this fascinating region, we need high resolution IR photometry and astrometry.

\begin{acknowledgements}
      This work has made use of data from the European Space Agency (ESA) mission {\it Gaia} (\url{https://www.cosmos.esa.int/gaia}), processed by the {\it Gaia} Data Processing and Analysis Consortium (DPAC, \url{https://www.cosmos.esa.int/web/gaia/dpac/consortium}). 
      Funding for the DPAC has been provided by national institutions, in particular the institutions participating in the {\it Gaia} Multilateral Agreement.
      This work uses data from the UKIDSS GPS catalog (version DR10PLUS) \citep{Lucas08}.
      The UKIDSS project is defined in \citet{Lawrence07}. UKIDSS uses the UKIRT Wide Field Camera \citep[WFCAM;][]{Casali07} and a photometric system described in \citet{Hewett06}. The pipeline processing and science archive are described in \citet{Hambly08}. 
      These work was supported by the Ministry of science and higher education of the Russian Federation by an agreement FEUZ-2023-0019.

\end{acknowledgements}

\appendix
\section{Determination of the interstellar absorption using the Q-method} \label{Ap}

\begin{figure}
      \centering
      \includegraphics[width=11cm]{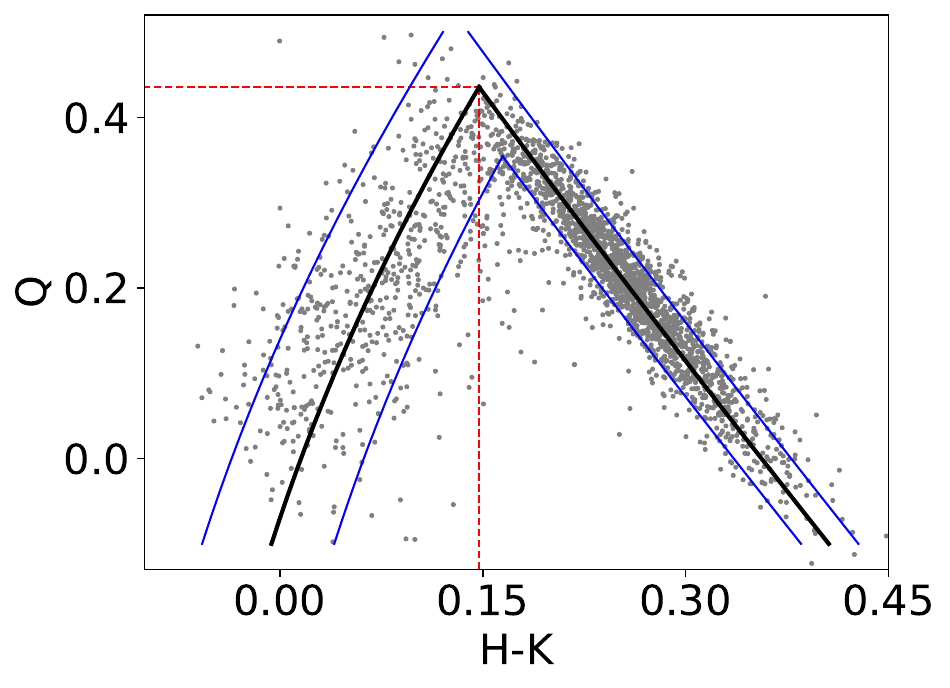}
      \caption{The zero-reddening sequence. 
      The Pleiades and Praesepe stars are shown by the gray dots, the non-reddened sequence described by the Eq.(\ref{ApEq_left}) and Eq.(\ref{ApEq_right}) is shown by solid black lines.
      Blue lines are the boundaries of the non-reddened sequence.
      The red dotted lines show the coordinates of the intersection point of the segments (see text for explanations).}
              \label{FigQ}%
\end{figure}

To estimate the individual absorption for stars, we used the Q-method first proposed by \citet{Johnson53}.
The essence of the method consists in comparing the position of the star relative to the non-reddened sequence on a Q-parameter versus color index diagram.
The parameter Q is extinction-free \citep{Johnson53}, and is constructed from a combination of the color indices and a ratio of the color excesses.
This ratio can be obtained from the extinction law.
For UKIDSS photometry we use the following expression for the Q parameter:

\begin{equation}\label{ApEq1}
Q=(J-H)-k_R(H-K),
\end{equation}

\noindent where $(J-H)$ and $(H-K)$ are color indices, $k_R=E(J-H)/E(H-K)=1.423$ \citep{Cardelli89,O'Donnell94}, $E(J-H)$ and $E(H-K)$ are color excesses for the corresponding color indices.

We plot the zero-reddening sequence using the stars of the Pleiades and Praesepe star clusters (from \citet {Lodieu19}).
Then, we approximate this sequence by the least squares on the plane (the parameter Q, the color index H-K) by two segments of a parabola (Fig.\ref{FigQ}).
Since the magnitudes of the stars in the Pleiades and Praesepe clusters are given in the 2MASS photometric system, they were transformed into the UKIDSS photometric standard using equations (6)–(8) of \citet{Hodgkin09}.
For fitting, we use the curve\_fit function of the \texttt{Python SciPy} library.
We obtained the following equations for the left and right segments, respectively:

\begin{equation}\label{ApEq_left} 
(H-K)=0.1573\cdot Q^2+0.2338\cdot Q+0.0157 ,
\end{equation}
\begin{equation}\label{ApEq_right} 
(H-K)=0.0337\cdot Q^2-0.4941\cdot Q+0.3561 .
\end{equation}

We calculate the color excess of a star as follows.
For a star, the parameter Q is calculated using the equation \ref{ApEq1}.
This Q value is then substituted into the equation of left (Eq. \ref{ApEq_left}) or right (Eq. \ref{ApEq_right}) segment to obtain the dereddened color index.
The segment selection method is described below.
Color excess is calculated as the difference between the original color index and the calculated absolute color index.
We convert the color excess obtained by the Q-method to the absorption in the selected band using the coefficients obtained by \citet{Cardelli89} and \citet{O'Donnell94} from the extinction curve: $A_J/A_V=0.28170$, $A_H/A_V=0.18251$ and $A_K/A_V=0.11281$, so $A_K\approx1.6185E(H-K)$ and $A_K\approx1.1373E(J-H)$.
These coefficients are extracted from the isochrone tables \citep{Bressan12}.
A limitation of the use of the Q-method is the range of Q values for which the zero-reddening  sequence is determined, in this work it is $Q\in[-0.10; 0.44]$.

In the plots `parameter Q~--~color index' for clusters of the G174+2.5 region, we see both stars which lie close to the zero-reddening sequence, and reddened stars located to the right of the absolute sequences.
We consider as unextincted those stars whose distance to the right segment (Eq. \ref{ApEq_right}) of the zero-reddening sequence along the H-K color index axis is less than 0.022. 
This value is the half-width of the right zero-reddening sequence.

We determine the width of each segment of the zero-reddening sequence individually by shifting the corresponding segment rightward and leftward along the color index axis.
The shift to the left or to the right was carried out until 10\% of the Pleiades and Praesepe stars belonging to this segment remained to the left or right of it, respectively.
Thus, between the boundaries of one segment there are 80\% of the stars belonging to this segment.
The distance between the shifted sequences of the segment was taken as the width of this segment.
The width of the left segment along the color index axis turns out to be  0.098, while that of the right segment is 0.044.
The boundaries of the un-reddened sequence obtained in this way are indicated in Fig. \ref{FigQ} by blue lines.

\begin{figure}
      \centering
      \includegraphics[width=10cm]{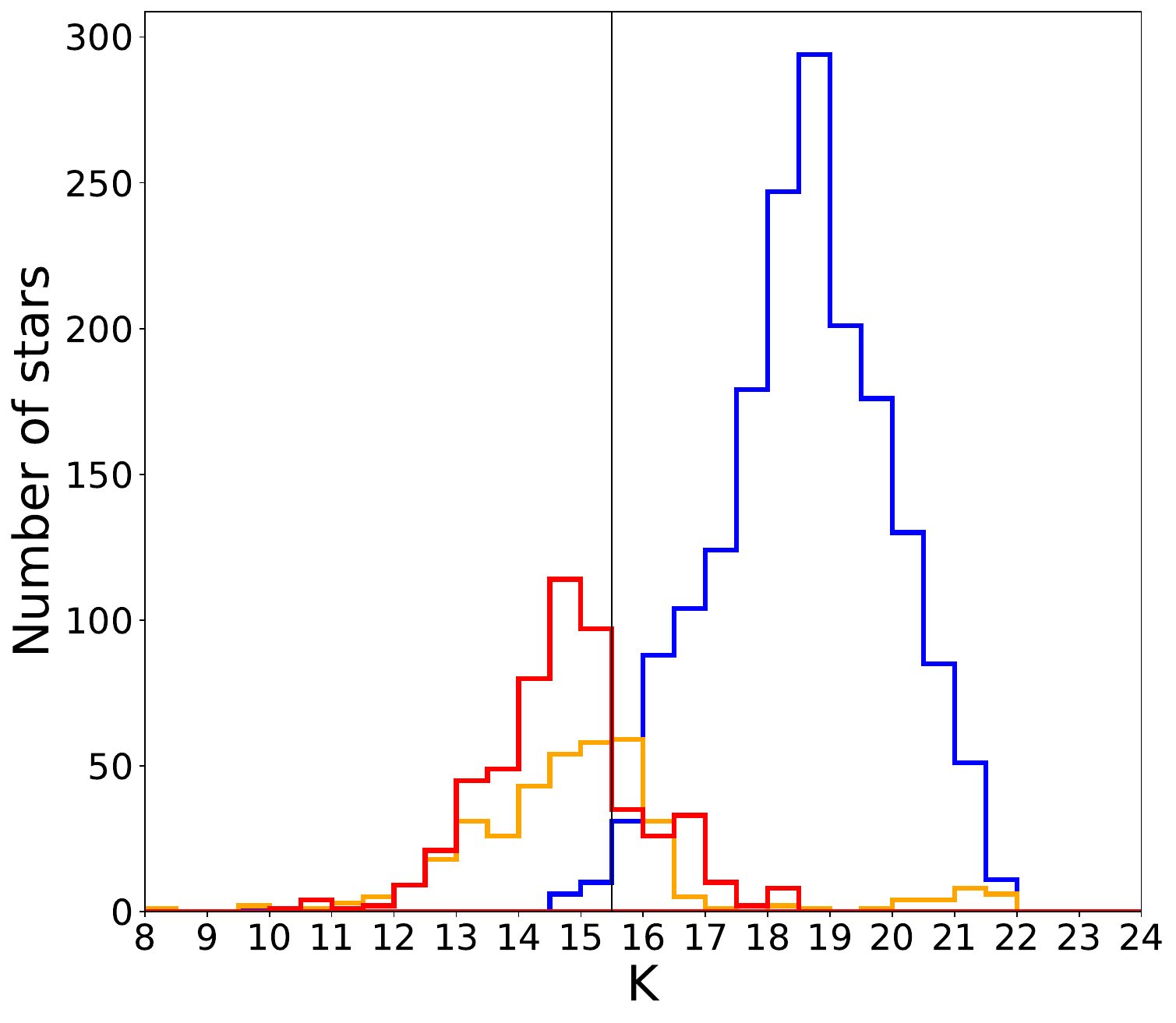}
      \caption{Combined luminosity functions of cluster 11 \textbf{(S235~A-B-C)} and the Pleiades.
      The luminosity function of the Pleiades stars belonging to the left segment is shown in orange, and the luminosity function of the right segment is shown in blue.
      The luminosity function of cluster 11 \textbf{(S235~A-B-C)} is shown in red.
      The black vertical line is the limiting value.}
              \label{FigLumFunc}%
\end{figure}

When using the Q-method, the main problem is choosing which segment of zero-reddening sequence a star belongs to, without knowing the spectral type of that star.
We tried to get around this issue in the following way.
In the Pleiades and Praesepe luminosity functions constructed using absolute magnitudes, the stars of the left and right segment occupy different magnitude ranges.
The stars on the left segment are brighter, while the stars on the right segment are dimmer.
These ranges overlap only in a small range of values (see Fig. \ref{FigLumFunc}).

A similar separation is observed for all adopted bands.

We define a magnitude limit at which the number of stars in the left segment still significantly exceeds the number of stars in the right segment.
The limit value is the same in both clusters when using the absolute magnitudes.
When constructing the luminosity function from apparent magnitudes, the initial limit value is shifted by the distance modulus of the corresponding cluster.
In this case, the sizes of the ranges of values in which the luminosity functions of the stars of the left and right segments are located, and the areas of their intersection, do not change.
Their relative position does not change either.
Therefore, we assume that the same separation of stars in the left and right segments exists for the clusters under study.
Then, by combining the luminosity function of the Pleiades and Praesepe with the luminosity function of the cluster, it is possible to approximately determine the limit value of the magnitude of our clusters.

For all stars in the cluster whose magnitude is less than this limiting value, the color excess is determined by the left segment (Eq. \ref{ApEq_left}) of the absolute sequence, for the rest~-- by the right (Eq. \ref{ApEq_right}).

The combination is carried out as follows.
First, we add the average extinction for the cluster under study, determined by the NICEST method, to the absolute magnitudes of the Pleiades and Praesepe stars.
Then, we shift these luminosity functions to the right until their left edge coincides with the left edge of the cluster's luminosity function.
From these shifted luminosity functions of the Pleiades and Praesepe, the limiting value for the cluster is determined.
An example of combining the luminosity function of cluster 11 \textbf{(S235~A-B-C)} and the Pleiades luminosity function is shown in Fig. \ref{FigLumFunc}.
We also note that when using limiting values defined for the same cluster but for different bands, the number of stars assigned to the left segment might vary.

When using this method of selecting the segment from which the absorption is determined, a discontinuity may occur in the photometric sequences on the CMD of clusters.
This is due to the fact that unreddened stars do not lie strictly on the absolute sequence, but occupy a region around it (limited by blue lines in Fig. \ref{FigQ}).
If there are no or few stars in the cluster with a Q parameter near 0.44, then, after correction for extinction, there will also be few stars on the CMD in the region with H-K color indices of about 0.15.
These values are the coordinates of the intersection point of the left and right segments of the absolute sequence (marked by the red dotted line in Fig. \ref{FigQ}).

\bibliography{Lit.bib}
\bibliographystyle{aasjournal}

\end{document}